%% file: template.tex
\documentclass[10pt,journal,compsoc]{IEEEtran}

\ifCLASSOPTIONcompsoc
  \usepackage[nocompress]{cite}
\else
  \usepackage{cite}
\fi

\newcommand{\arxiv}

\input{contents/preamble.tex} %
\usepackage{ragged2e}

\ifCLASSINFOpdf
\else
\fi

\usepackage{array}

\begin{document}
\title{\addm{Interactive Volume Visualization via Multi-Resolution Hash Encoding based \\ Neural Representation}}

\author{Qi~Wu,
        David~Bauer, 
        Michael~J.~Doyle,
        and~Kwan-Liu~Ma%
\IEEEcompsocitemizethanks{%
\IEEEcompsocthanksitem Qi Wu, David Bauer and Kwan-Liu Ma are with\\%
University of California - Davis, USA.\protect\\%
E-mail: \{qadwu\,$|$\,davbauer\,$|$\,klma\}@ucdavis.edu
\IEEEcompsocthanksitem Michael J. Doyle is with Intel Corporation, USA.\protect\\
E-mail: michael1.doyle@intel.com%
}%
}

\ifdefined\arxiv\else

\ifCLASSOPTIONpeerreview
    \markboth{IEEE Transactions on Visualization and Computer Graphics}%
    {Instant Neural Representation for Interactive Volume Rendering}
\else
    \markboth{IEEE Transactions on Visualization and Computer Graphics}%
    {Wu \MakeLowercase{\textit{et al.}}: Instant Neural Representation for Interactive Volume Rendering}
\fi

\fi

\IEEEtitleabstractindextext{%
\begin{abstract}\justifying
\input{contents/abstract.tex}
\end{abstract}%
\ifCLASSOPTIONpeerreview
\else
\fi
}

\maketitle

\IEEEdisplaynontitleabstractindextext

\IEEEpeerreviewmaketitle

\IEEEraisesectionheading{\section{Introduction}\label{sec:introduction}}

\input{contents/1.intro.tex}
\input{contents/2.related.tex}

\input{contents/3.method.tex}

\input{contents/4.features.tex}

\input{contents/5.rendering.tex}

\input{contents/3.ablation}

\input{contents/6.evaluation.tex}
\input{contents/conclusion.tex}

\ifCLASSOPTIONpeerreview
\else

\ifCLASSOPTIONcompsoc
  \section*{Acknowledgments}
\else
  \section*{Acknowledgment}
\fi

\input{contents/acknowledgments.tex}

\fi

\ifCLASSOPTIONcaptionsoff
  \newpage
\fi

\bibliographystyle{IEEEtran}
\bibliography{IEEEabrv,template}

\ifCLASSOPTIONpeerreview
\else

\ifdefined\arxiv\else

\begin{IEEEbiography}{Qi~Wu}
Biography text here.
\end{IEEEbiography}

\begin{IEEEbiography}{David~Bauer}
Biography text here.
\end{IEEEbiography}

\begin{IEEEbiography}{Michael~J.~Doyle}
Biography text here.
\end{IEEEbiography}

\begin{IEEEbiography}{Kwan-Liu~Ma}
Biography text here.
\end{IEEEbiography}

\fi

\fi

\ifdefined\arxiv
\newpage\appendices
\input{contents/7.supplementary.tex}
\fi

\end{document}

%% file: contents/preamble.tex
\usepackage{xspace}
\usepackage[dvipsnames]{xcolor}
\usepackage{soul}
\usepackage{comment}
\usepackage{multirow}
\usepackage{listings}
\usepackage{amsmath}
\usepackage{amssymb}
\usepackage{adjustbox}
\usepackage{enumitem}
\usepackage{hyperref}
\usepackage{cleveref}
\usepackage[caption=false, font=footnotesize]{subfig}
\usepackage[normalem]{ulem}
\usepackage{multicol}
\usepackage{makecell}
\usepackage{tabu}
\usepackage{booktabs, siunitx}
\usepackage{algpseudocode}
\usepackage{algorithm}

\definecolor{codeGreen}{rgb}{0,0.6,0}
\definecolor{codeBlue}{rgb}{0,0,1}
\definecolor{codeRed}{rgb}{0.65,0.11,0.36}
\definecolor{codeGray}{rgb}{0.6,0.6,0.6}
\definecolor{codeMauve}{rgb}{0.58,0,0.82}
\definecolor{codeCyan}{rgb}{0,0.52,0.70}

\newcommand{\ie}{\MakeLowercase{i.e.,}\xspace}

\newcommand{\etal}{\MakeLowercase{et al.}\xspace}

\newcommand{\tcnn}{Tiny-CUDA-NN\xspace}
\newcommand{\neurcomp}{\textit{neurcomp}\xspace}
\newcommand{\tthresh}{\textit{tthresh}\xspace}
\newcommand{\fvsrn}{\textit{fV-SRN}\xspace}

\newcommand{\dataset}[1]{\textbf{#1}}

\newcommand{\best}[1]{\underline{#1}}
\newcommand{\last}[1]{#1}

\newcommand{\addm}[1]{{#1}}
\newcommand{\add}[1]{{#1}}

\newcommand{\michaeledits}

\def\lstfontsize{\fontsize{7}{7}\selectfont}

%% file: contents/abstract.tex
Implicit neural networks have demonstrated immense potential in compressing volume data for visualization. However, despite their advantages, the high costs of training and inference have thus far limited their application to offline data processing and non-interactive rendering. In this paper, we present a novel solution that leverages modern GPU tensor cores, a well-implemented CUDA machine learning framework, an optimized global-illumination-capable volume rendering algorithm, and a suitable acceleration data structure to enable real-time direct ray tracing of volumetric neural representations. Our approach produces high-fidelity neural representations with a peak signal-to-noise ratio (PSNR) exceeding 30 dB, while reducing their size by up to three orders of magnitude. Remarkably, we show that the entire training step can fit within a rendering loop, bypassing the need for pre-training. Additionally, we introduce an efficient out-of-core training strategy to support extreme-scale volume data, making it possible for our volumetric neural representation training to scale up to terascale on a workstation with an NVIDIA RTX 3090 GPU. Our method significantly outperforms state-of-the-art techniques in terms of training time, reconstruction quality, and rendering performance, making it an ideal choice for applications where fast and accurate visualization of large-scale volume data is paramount.

%% file: contents/1.intro.tex
\IEEEPARstart{M}{}odern simulations and experiments can produce massive amounts of high-fidelity volume data, which provide the necessary details for understanding complex scientific processes. These data are challenging to visualize interactively due to their sheer size. Neural networks have recently shown promising potentials for compactly and implicitly parameterizing continuous volumetric fields~\cite{sitzmann2020implicit}. 
In the space of scientific visualization, this method has been applied to represent volume data by Lu~\etal~\cite{lu2021compressive}. 
Their approach directly approximates the mapping from spatial coordinates to volume values using a multilayer perceptron (MLP). The trained MLP is then considered a compressed version of the original data. This representation is efficient because the memory footprint of a neural network is often orders of magnitude smaller than the original data. Sampling the representation is also flexible, as one can arbitrarily query volume values without explicit decompression and interpolation. However, complex MLPs (with hundreds of thousands of trainable parameters) are often needed to capture high-frequency details in the volume. The use of complex MLPs makes training and rendering prohibitively expensive. Furthermore, this method has not been tested on large-scale volume data generated by state-of-the-art simulations and experiments.

Recent progress in Neural Radiance Field (NeRF) has shown that smaller MLPs with high-dimensional trainable input encodings can accurately learn complex volumetric fields. This method allows us to capture high-frequency details in the data with relatively small models. In addition, by moving most of the network parameters to the encoding layer, training can be accelerated since each backward propagation process only updates a small number of parameters~\cite{muller2022instant, takikawa2021neural, martel2021acorn}. Advancements in machine learning hardware and software can further accelerate training and inference.
\add{In scientific visualization, Weiss~\etal's \fvsrn~\cite{weiss2021fast} adopts this approach, organizing encoding parameters into a single-resolution dense grid (referred to as the \emph{latent grid} in their report). They also developed a custom CUDA kernel to accelerate network inference using tensor cores. However, their CUDA kernel does not accelerate network training, and the use of dense-grid encoding can result in a substantial memory footprint for large-scale volume data.}

In this work, we build on the successes of these techniques and introduce them to the scientific visualization domain. Our work \add{also} models volume data using neural representations in the form of compact MLPs, \add{but} leverages the multi-resolution hash grid encoding method recently proposed by M\"uller~\etal~\cite{muller2022instant} to capture high-frequency details in the data. Then, we use the tensor-core-accelerated \tcnn~\cite{tiny-cuda-nn} machine learning framework and a pure CUDA implementation to maximize training speeds.

Although these training improvements are highly effective, significant gaps remain if we are to meet the typical performance and scalability requirements of volume visualization applications. Specifically, we must provide methods for the interactive rendering of these neural representations and methods for handling large-scale volume data.

\begin{figure*}[tb]
    \centering
    \includegraphics[width=\linewidth]{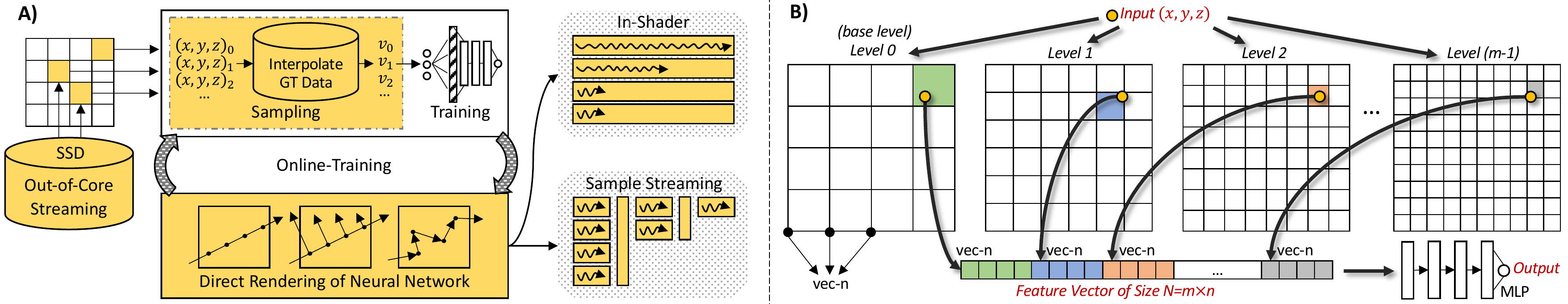}
    \vspace{-1em}
    \caption{A) An overview of our work. The sampling step randomly and uniformly generates sample using the ground truth (GT) data. The ground truth data can be loaded via out-of-core streaming. The training step optimizes the neural network. The rendering step renders the neural network via in-shader or sample-streaming methods. Our approach accommodates both pre-training and online-training. Our novel contributions are highlighted in yellow. B) The architecture of our neural network with the multi-resolution hash grid encoding method~\cite{muller2022instant}.}
    \label{vnr:fig:network-structure}
\end{figure*}

To enable the interactive rendering of volumetric neural representations, we develop a \emph{sample streaming} rendering technique. This method iteratively interrupts the ray tracing kernel and streams out sample coordinates for network inference. 
We compare this \emph{sample streaming} method to fully-customized \add{CUDA rendering kernels that perform network inference \emph{in shader}, similar to the method used by Weiss~\etal~\cite{weiss2021fast}.}
Our \emph{sample streaming} method is about 3-8$\times$ faster than \emph{in-shader}. Additionally, we introduce a macro-cell acceleration structure~\cite{novak2014residual, szirmay2011free} for rendering our neural representations. We demonstrate that this data structure is particularly suitable for accelerating this type of neural rendering and can bring up to a 40$\times$ speedup. As a result, our \emph{sample streaming} method can deliver real-time performance (up to 200fps) and advanced global illumination via ray marching shadows and unbiased path tracing.

To provide a complete rendering solution for large-scale volume data, we develop out-of-core data streaming techniques.
Training a neural network typically requires loading the entire dataset to GPU or system memory, which is infeasible for large-scale data. A simple workaround would use virtual memory and file mapping. However, this approach is unsuitable for training volumetric neural representations, where data values are accessed randomly. For large-scale data, such a data access pattern can lead to enormous page swapping, leading to poor I/O performance. Our method maintains an ever-changing random subset of the data in memory and decouples data streaming from the network training. Thus, we can regularize the data access pattern. We demonstrate that this method is more than $10\times$ faster than the virtual memory implementation. We also demonstrate that it is feasible to fit a network training step inside our rendering loop and achieve online training---even for extreme-scale data.
\addm{We release the source code of our implementation at \url{https://github.com/VIDILabs/instantvnr}.}

An overview of our work is provided in \Cref{vnr:fig:network-structure}A. We also summarize our contributions as follows:
\begin{itemize}
\item \addm{A compressive volumetric neural representation that utilizes multi-resolution hash encoding~\cite{muller2022instant}, providing nearly instantaneous training performance.}
\item A fast \emph{sample streaming} algorithm to directly ray trace volumetric neural representations at up to 60fps with advanced illuminations, significantly outperforming the state-of-the-art.
\item An efficient \emph{out-of-core sampling} scheme that allows neural representation training to scale up to very large-scale on a single-GPU workstation for the first time.
\item \add{The demonstration of} interactive online training for volumetric neural representations.
\end{itemize}

%% file: contents/2.related.tex
\section{Related Work}

Because we focus on creating compact neural representations for volume data, we first give an overview of related deep-learning methods for volume compression. Then, as our work uses multi-resolution hash grid encoding, we also provide the background of input encoding techniques.

\textbf{Deep Learning for Volume Compression.}
High-quality volume visualization can be challenging, in part because handling large-scale, high-resolution volume data can be difficult. Therefore, deep learning techniques have been explored to compress volume data. 
Earlier work by Jain~\etal~\cite{jain2017compressed} presented an encoder-decoder network to compress a high-resolution volume. %
Wurster~\etal~\cite{wurster2021deep} later used a generative adversarial networks (GANs) hierarchy to complete the same task. 
Super-resolution neural networks can directly work with a low-resolution volume and upscale it when necessary. This technique can be helpful when it is too expensive to store the data in high resolution~\cite{zhou2017volume, han2020ssr, guo2020ssr}, or for all timesteps~\cite{han2019tsr,han2021stnet}, or both~\cite{han2021stnet}. 
This technique can also accelerate simulations by allowing them to work on lower-resolution grids~\cite{xie2018tempogan}. %
Lu~\etal~\cite{lu2021compressive} explored using implicit neural representations, as mentioned previously. %
However, their method requires a time-consuming training process for every volume data. 
The concurrent work performed by Weiss~\etal~\cite{weiss2021fast} improves this work by employing dense-grid encoding~\cite{takikawa2021neural} for faster training and GPU tensor core acceleration for faster inference. 
Doyub~\etal~\cite{kim2022neuralvdb} recently integrated implicit neural representation into the OpenVDB framework for handling high-resolution sparse volumes. 
Our work improves Lu~\etal's method but focuses on achieving interactive rendering algorithms and supporting large-scale volume data.

\textbf{Out-of-Core Data Streaming}. Volume rendering with out-of-core data streaming is a well-established area of research. Early work by LaMar \etal~\cite{lamar1999multiresolution} and Weiler \etal~\cite{zimmermann2000level} combined data streaming and texture LoD for large-scale volume rendering. Gobbetti \etal~\cite{gobbetti2008single} later introduced a similar algorithm using octrees, while GigaVoxels~\cite{crassin2009gigavoxels} optimized this approach with ray-guided streaming for rendering voxel surfaces. CERA-TVR~\cite{engel2011cera} generalized GigaVoxels to general-purpose volume visualization. Tuvok~\cite{fogal2010tuvok} coupled data streaming with progressive rendering and introduced a caching mechanism that combines the most recently used and least recently used strategies. Hadwiger~\etal~\cite{hadwiger2012interactive} presented the multi-resolution page-directory technique for streaming terascale imaging volumes from GPU, and Sarton \etal~\cite{sarton2019interactive} extended this approach to all regular volumes. Recently, Wu~\etal~\cite{wu2022flexible} introduced a flexible data structure to stream and cache unstructured and adaptive mesh refinement (AMR) volumes, while Zellmann~\etal~\cite{zellmann2022design} enabled interactive visualization for terascale time-varying AMR data using low-latency hardware APIs.

\textbf{Network Input Encoding.}
We apply the idea of positional encoding that first maps input coordinates to a higher-dimensional space in our neural network, as it allows MLPs to capture high-frequency, local details better. One-hot encoding~\cite{harris2010digital} and the kernel trick~\cite{theodoridis2006pattern} are early examples of techniques that make complex data arrangements linearly separable. 
In deep learning, positional encodings are helpful for recurrent networks~\cite{gehring2017convolutional} and transformers~\cite{vaswani2017attention}. In particular, they encode scalar positions as a sequence of sine and cosine functions. NeRF work and many others~\cite{tancik2020fourier, barron2021mip, muller2019neural} use similar encoding methods, which are often referred to as \emph{frequency encodings}.
More recent works introduced \emph{parametric encodings} via additional data structures such as dense grids~\cite{martel2021acorn}, sparse grids~\cite{hadadan2021neural}, octrees~\cite{takikawa2021neural}, or multi-resolution hash tables~\cite{muller2022instant}. 
By putting additional trainable parameters into the encoding layer, the neural network size can be reduced. 
Therefore, neural networks with such encoding methods can typically converge much faster while maintaining approximately the same accuracy. 
In this work, we adopt the multi-resolution hash grid method proposed by M{\"u}ller~\etal~\cite{muller2022instant} because of its excellent performance in training MLPs.

%% file: contents/3.method.tex
\section{Network Design}

Volume data in scientific visualization essentially can be written as a function $\Phi$ that maps a spatial location $(x,y,z)$ to a value vector $\textbf{v}$ which represents the data value at that spatial location
$\Phi:$ $\mathbb{R}^3 \rightarrow \mathbb{R}^D,~(x,y,z) \mapsto \Phi(x,y,z) = \textbf{v}$.
Such a volumetric function is typically generated by simulations or measurements and then discretized, sampled, and stored.
In this work, we focus on scalar field volume data ($D$=$1$) and employ an implicit neural network to model the volumetric function $\Phi$ by optimizing it directly using sample coordinates and \add{data} values\add{~before transfer function classifications, as suggested by Lu et al.\cite{lu2021compressive}}.
Since the network is defined over the continuous domain $\mathbb{R}^3$, it can directly calculate data values at an arbitrary spatial resolution and avoid explicit interpolations. 
Moreover, since the network processes each input position independently, data values can also be sampled on demand.
Additionally, because the neural network approximates the volume data analytically, the number of neurons needed to represent the volume data faithfully does not increase linearly as the data resolution increases, promising greater scalability potentially.
Finally, although only scalar volumetric fields are studied in this work, we believe that the same method can be easily extended to multivariate cases.
Weiss~\etal~\cite{weiss2021fast} have reported findings in this direction by jointly training gradients and curvatures.

\subsection{Network Implementation}\label{vnr:sec:network-impl}

We implemented our neural representation in CUDA and C++ using the \tcnn machine learning framework~\cite{tiny-cuda-nn}. Unless specified otherwise, the ``fully-fused-MLP'' is used in this paper. By fitting inputs into the shared memory and weights into registers, this MLP implementation can more efficiently utilize GPU tensor cores.
Additionally, as indicated by Mildenhall~\etal~\cite{mildenhall2020nerf} and Tancik~\etal~\cite{tancik2020fourier}, an input encoder that first converts coordinates to high-dimensional feature vectors is needed. 
Such an encoder allows the network to capture high-frequency features in the data more efficiently.
We find that  the multi-resolution hash grid encoding method~\cite{muller2022instant}, initially proposed for NeRF and other computer graphics applications, is very effective for scientific volume data. Therefore we employ this method in our implementation.

As illustrated in \Cref{vnr:fig:network-structure}B, this hash grid encoding method defines $m$ logical 3D grids over the domain. 
Each grid is referred to as a level and is constructed using a hash table of size $T$. Each table entry stores $n$ trainable parameters (also referred to as features).
Thus, each logical grid point can be mapped to a feature vector of size $n$ via the associated hash table.
For a given input coordinate, a feature vector of size $n$ can be calculated for each level by interpolating nearby grid features. The final feature vector of size $N$$=$$n$$\times$$m$ can then be constructed via concatenation. 
We use this feature vector as the input for the MLP.
This input encoding method is key to how our neural representation can support high-fidelity  volume rendering.

Throughout this paper, we use ReLU activation functions for all MLP layers. We use the Adam optimizer~\cite{kingma2014adam} with an exponentially decaying learning rate to optimize networks. We use L1 loss unless specified otherwise. For each training step, we take precisely 65536 training samples.
Through an ablation study (\Cref{vnr:sec:ablation}), we observe that varying the hash table size, the number of hash encoding layers, the number of encoding features per layer, and the number of hidden layers in the MLP can all have significant impact to the volume reconstruction quality, rendering quality, and compression ratio. Thus, we apply different network parameters to different datasets to accommodate their different characristics.

%% file: contents/4.features.tex
\section{Training}

We describe our training in terms of training steps. 
Note that a training step is different from an epoch. The latter indicates one complete pass of the training set through the algorithm. In our context, the training set is an infinite and continuous domain; thus, the former term is used. A single step is conducted by first generating a batch of random coordinates (batch size = 65536) within the normalized domain $[0,1)^3$. 
Then we reconstruct the corresponding scalar value for each coordinate using the ground truth. The returned scalar value is normalized to $[0, 1]$.
This paper refers to this step as the ``sampling'' step. 
Since all volumetric function values are generated independently, the sampling step is embarrassingly parallelizable.
We use a normalized domain and range here because we observe that coordinates and values with high dynamic range can make the network very unstable, if not impossible, to optimize.
Once the sampling is done, the input and the ground truth data are loaded into the GPU memory (if they are not already there) and then used to optimize our volumetric neural representation using the \tcnn's C++ API.

When the target volume is larger than the GPU memory, 
we can offload the sampling process to the CPU and reconstruct training samples with the help of multi-threading and SIMD vectorization.
In our case, we implement our CPU-based sampling process using the OpenVKL~\cite{OpenVKL} volume computation kernel library.
As OpenVKL already supports a variety of volume structures, we can also apply our volumetric neural representations beyond dense grids. Our preliminary study in this direction also shows promising results. However, as this is mainly outside the scope of this paper, we leave the study and evaluation of sparse volumes to future works.

\begin{figure}[tb]
    \centering
    \includegraphics[width=0.8\linewidth]{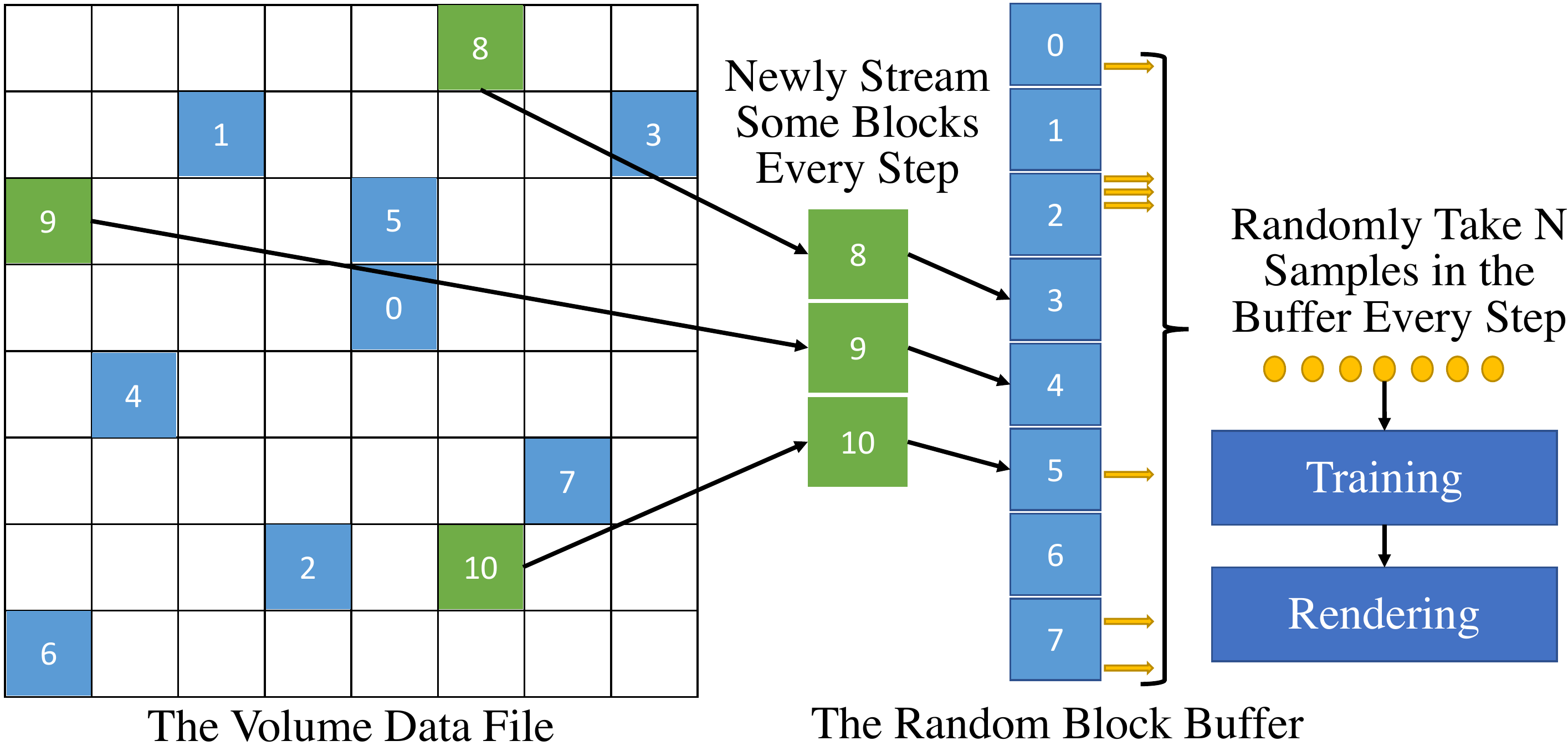}
    \vspace{-0.1in}
    \caption{Our out-of-core sampling method works by maintaining a random block buffer in the system memory. We sample and asynchronously update this buffer every frame. Then, we transfer sampled results to GPU for training.}
    \label{vnr:fig:out-of-core-training}
    \vspace{-0.15in}
\end{figure}

\subsection{Out-of-Core Sampling}\label{vnr:sec:training-ofc}

Modern scientific simulations and experiments often produce high-resolution datasets. These datasets often cannot be loaded into even the system memory. Thus, the sampling methods mentioned previously will not work.
Using virtual memory and file mapping can be a workaround. However, as we draw training samples randomly, this workaround would lead to enormous page swappings and produce significant latencies.
We develop a novel way to handle this situation, which we call \emph{out-of-core sampling}.
As illustrated by~\Cref{vnr:fig:out-of-core-training}, this method maintains a buffer of $R$ randomly selected 3D blocks in the system memory. 
Each 3D block in the buffer is roughly 64KB and can be randomly mapped to a 3D region in the volume data domain via asynchronous data streaming.
We use blocks of roughly 64KB because it yields a good balance between streaming performance and data scalability~\cite{wu2022flexible}. The actual shape of a 3D block does not significantly impact the training performance and training quality.
The first sampling step will properly initialize all the blocks  to ``warm up'' the random buffer.
Then we refresh only $S$ blocks (obviously $S \leq R$) in every subsequent sampling step.

To take samples, instead of uniformly sampling random coordinates, we first randomly select a block within the random buffer. Then randomly select a voxel within the selected block and calculate the corresponding voxel index.
Finally, we jitter the voxel center by half a voxel and use the jittered $(x_{c} + X, y_{c} + Y, z_{c} + Z~|~ X,Y,Z \sim \mathcal{U}_{[-0.5,0.5]})$ coordinate as the sample.
The corresponding sample value is then interpolated using values of the voxel and its neighbors.
When trilinear interpolation is enabled, each block will also include a layer of additional ghost voxels surrounding the block that are necessary for performing the interpolation.

Although we implement the mapping via asynchronous I/O, we still perform a synchronization step before the start of each sampling process. This synchronization step allows us to simplify our sampling implementation at the cost of lower I/O bandwidth. 
We leave the fully asynchronous implementation to future works.
Our implementation uses an NVMe SSD as the data storage device because it offers superior random read bandwidth. 
We default $S=1024$ to maintain a relatively good sampling performance and a reasonably large $R=65536$ for a good initial guess of the full-resolution data.
We evaluated the effectiveness of out-of-core sampling and the choice of $S$ and $R$ in \Cref{vnr:eval:sampling}.

\subsection{Online Training and Rendering}

With the help of \tcnn, hash grid encoding, and our optimized sampling implementation, it is now possible to fit the entire training step inside the rendering loop while still achieving interactive visualization.
Such an online-training capability allows users to interactively and visually examine the neural representation optimization process. It can also potentially accelerate the debugging and hyperparameter tuning process in practice.
Online training also provides users with a new and elegant way to quickly preview extremely large datasets using a computer with limited RAM and VRAM.
For example, the double-precision channel flow DNS~\cite{dns} volume is around 950GB on disk.
Traditionally, out-of-core streaming and progressive rendering techniques~\cite{hadwiger2012interactive, wu2022flexible} would be required to handle such data on a single machine.
However, with online training, not only can the output of training be saved and reused in the future, but there is also no need to use a specialized progressive renderer. 
Since the training and rendering are entirely decoupled, the complexity of handling large-scale data is reduced.

%% file: contents/5.rendering.tex
\section{Rendering}

Decoding the volumetric neural representation back to a dense grid is a na\"ive way to achieve interactive volume visualization. Such an approach would limit the ability to render large-scale volume data directly. The decoding process can be done progressively in every frame for online training. However, it can lead to noticeable rendering artifacts, especially in the early stage of training. Therefore, we have developed two novel techniques to overcome these drawbacks and render volumetric neural representations. For each technique, we have provided the implementations of three rendering algorithms: ray-marching with the emission-absorption model, ray-marching with single-shot heuristic shadows~\cite{Brownlee2021} (only shading at the point of highest contribution), and volumetric path-tracing.

\begin{figure}[tb]
\vspace{-0.75em}
\begin{lstlisting}
template<RayTypeEnum TYPE, typename Params> 
void streaming_renderer(Params& params, int n_rays) {
  ^raygen_kernel<TYPE>^<<<...>>>(params, n_rays);
  while (~compaction~(&n_rays)) {
    ^coordinate_kernel<TYPE>^<<<...>>>(params, n_rays);
    batch_inference(params.net, params, n_rays);
    ^shading_kernel<TYPE>^<<<...>>>(params, n_rays);
  }
}

// ray marching with absorption-emission model
void raymarching_simple(Params& params, int n_rays) {
  streaming_renderer<CAMERA_RAY>(params, n_rays);
}

// ray marching with single-shot heuristic shadow
void raymarching_shadow(Params& params, int n_rays) {
  streaming_renderer<CAMERA_RAY>(params, n_rays);
  // using the highest contribution sample's shadow 
  // to attenuate the entire primary ray 
  streaming_renderer<SHADOW_RAY>(params, n_rays);
}
\end{lstlisting}
\vspace{-1.2em}
\caption{Ray marching algorithms using sample-streaming.}%
\label{vnr:fig:sample-streaming-code}
\vspace{-1em}
\end{figure}

\subsection{In-Shader Inference}\label{vnr:sec:rendering-in-shader}

We develop a customized routine using native CUDA to directly infer the neural network inside a shader program, as inspired by M{\"u}ller~\etal~\cite{muller2021real}. This routine eliminates global GPU memory accesses, fuses the hash grid encoding calculation into the MLP, and manages network inputs and outputs locally within each threadblock via shared memory.
Each CUDA warp computes the multiplication between 16 neurons and 16 MLP inputs, which is required to utilize tensor cores for hardware acceleration. 
We also provide a templated CUDA device API for launching arbitrary rendering kernels with correct threadblock configurations and shared memory sizes. This method is also concurrently developed by Weiss~\etal~\cite{weiss2021fast}. However, our implementation does not require the MLP width to be identical to the output size of the encoding layer. Above this, it performs better in general (as shown in \Cref{vnr:tab:quality}) and is fully compatible with networks produced using the \tcnn framework. This rendering approach requires only small changes in the rendering algorithms, thus can be easily adopted in many complicated systems.  However, the amount of low-level optimization being used, we discover that significant performance gains can be obtained by completely overhauling the rendering algorithm, as we explain in the following section.

\subsection{Sample Streaming}\label{vnr:sec:rendering-sample-streaming}

The machine learning framework we use is optimized for batch training and inference, meaning that a relatively large number of inputs need to be pre-generated.
Then the neural network infers all the inputs at the same time. 
This execution model can achieve higher GPU performance because it reduces control flow divergence, thus improving thread utilization. It can also effectively hide high-latency memory accesses as there are enough data for the processor to work on while waiting for a memory request. 
However, such an execution model can make the rendering algorithm more complex.
To take full advantage of the batch inference execution model, we need to fundamentally change the rendering algorithm.
Our implementations share the same spirit as wavefront path tracers~\cite{laine2013megakernels} except that we have to solve a very different challenge: volume densities are unknown in advance and expensive to compute.

\begin{algorithm}[tb]
\caption{\label{alg:woodcock}Woodcock tracking with null collisions.}
\begin{algorithmic}
\Function{Woodcock~}{$\mu_{\max}$, $t_{\min}$, $t_{\max}$}
    \State $t = t_{\min}$
    \While{$t < t_{\max}$} \Comment{ray inside the volume}
        \State $\zeta,\xi \sim\mathcal{U}_{[0,1]}$       
        \State $t = t - \frac{\log{(1-\zeta)}}{\mu_{\max}}$
        \If{$\xi < \frac{\mu(t)}{\mu_{\max}}$}
            \State \Return{$t$} \Comment{real collision $\Rightarrow$ exit loop}
        \EndIf{}             \Comment{or null collision $\Rightarrow$ continue}
    \EndWhile{}
    \State \Return{INVALID} \Comment{report no real collision}
\EndFunction
\end{algorithmic}
\end{algorithm}

\begin{figure}[tb]
\vspace{-0.75em}
\begin{lstlisting}
// generate a sample coordinate
bool sample(Ray& ray) {
  float t;
  if (ray.has_hit(t)) {
    ray.next_coordinate = ray.at(t); return true;
  }
  if (ray.shadow) {
    // compute direct lighting ...
    ray.shadow = false; // switch to scattering ray ...
    // re-generate a coordinate ... 
    if (ray.has_hit(t)) {
      ray.next_coordinate = ray.at(t); return true;
    }
  }
  return false;
}

bool shade(Params& params, Ray& ray) {
  if (ray.is_null_collision()) return true;
  if (ray.shadow) {
    ray.shadow = false; // switch to scattering ray ...
  }
  else {
    if (russian_roulette(ray)) return false;
    ray.throughput *= phase_function(params, ray);
    ray.org = ray.next_coordinate; // advance ray
    ray.shadow = true; // switch to shadow ray ...
  }
  return ray.intersect_volume();
}

template<>
void raygen_kernel<PT_RAY>(Params& params, int n_rays) {
  Ray ray = // initialize the ray ...
  bool valid = sample(ray);
  if (valid) save_ray(params, ray);
  else accumulate_pixel_color(params, ray);
}

template<>
void coordinate_kernel<PT_RAY>(Params&, int) {}

template<>
void shade_kernel<PT_RAY>(Params& params, int n_rays) {
  Ray ray = // load the ray from global memory ...
  bool valid = shade(params, ray) && sample(ray);
  if (valid) save_ray(params, ray);
  else accumulate_pixel_color(params, ray);
}

void pathtracing(Params& params, int n_rays) {
  streaming_renderer<PT_RAY>(params, n_ray);
}
\end{lstlisting}
\vspace{-1.2em}
\caption{The path tracing algorithm using sample-streaming.}
\label{vnr:fig:sample-streaming-pt-code}
\vspace{-1.2em}
\end{figure}

\subsubsection{Ray Marching}\label{vnr:sec:sample-streaming-rm}

We start by focusing on ray marching and splitting the algorithm into three smaller kernels: 
the ray-generation kernel, the coordinate-computation kernel, and the shading kernel
(as highlighted in red in \Cref{vnr:fig:sample-streaming-code}).
The ray-generation kernel initializes all the primary rays and intersects them with the volume. 
If no intersection is found, the ray is invalidated. 
We remove invalid rays via stream compaction (green in \Cref{vnr:fig:sample-streaming-code}). 
This operation is also performed for subsequent iterations in the render pass.
Within the loop, the next $K$ sample coordinates for each ray are streamed out for network inference. 
Then, inferred sample values are retrieved by another kernel for shading. 
Finally, the loop terminates if all the rays are invalid.

When the number of samples generated by each iteration $K$ is greater than 1, the number of iterations needed to complete a frame can be reduced. This optimization reduces CUDA kernel launch overheads.
Additionally, because exited rays are removed at the end of each iteration, the CUDA kernel size within each iteration will decrease as the loop iterates.
Increasing $K$, small tailing kernels can be batched up to reduce the launch overhead further.
However, if $K$ is too large, the number of samples being computed per iteration can be more than necessary because many rays might have exited earlier. These unused samples can only be discarded, and the time spent to compute these samples is wasted.
Therefore, in order to achieve optimal performance, $K$ needs to be tuned. 
Our tuning result suggests that %
the algorithm performs best when $K$ is around 8, which is the value we used in this paper.
The tuning experiment is described in the appendix due to space limitations.
Notably, this optimization currently only applies to ray marching algorithms.
This optimization is referred to as \textit{batched ray marching} in this paper.

\subsubsection{Path Tracing}

In line with current scientific visualization trends, we have incorporated volumetric path tracing in our study. Unlike ray marching, this method employs distance sampling techniques to calculate the distance a ray can travel within a non-uniform volume before undergoing interactions. Our work adopts the Woodcock tracking~\cite{woodcock:1965} algorithm, owing to its simplicity (refer to \Cref{alg:woodcock}). This algorithm utilizes fictitious particles as control variates to homogenize the non-uniform media. It utilizes the mixing ratio between fictitious particles and real particles at the interaction point to probabilistically determine whether the interaction is real or not. A real interaction triggers direct lighting and scattering, while a null collision results in the ray continuing to travel in the same direction. To estimate direct lighting, multiple passes of Woodcock tracking towards light sources are required. The present multiple Woodcock tracking passes means that the volume is sampled within multiple nested loops, which significantly complicates the sample streaming implementation. As a result, we have to completely restructure the path tracing algorithm to reduce the number of streamouts per frame.

Similar to our ray marching approach, our path tracing implementation creates multiple smaller kernels, as illustrated in \Cref{vnr:fig:sample-streaming-pt-code}. However, unlike ray marching, the ray-generation and shading kernels absorb the sample computation kernel. This is because the ray marching algorithm can potentially take multiple samples per iteration; thus it is more efficient for ray marching to compute sample coordinates separately after compaction, and organize samples in a structure-of-arrays fashion to facilitate coalesced global memory accesses.

Our path tracing implementation maintains a single active ray and alternates its role between the \emph{scattering} and \emph{shadow} states. When a real collision is found using Woodcock tracking, a shadow ray is constructed, and the same tracking algorithm is used to sample the light source to estimate transmittance~\cite{pharr2016physically}. Subsequently, the ray generates a new scattering direction, transitions to the scattering state, and continues propagation within the volume using Woodcock tracking. Ray termination occurs when the ray exits the volume or based on Russian Roulette~\cite{pharr2016physically}. The path tracing process primarily comprises two functions, as depicted in \Cref{vnr:fig:sample-streaming-pt-code}. The \texttt{sample} function computes the next sample coordinate for the current ray and potentially converts a shadow ray to a scattering ray, whereas the \texttt{shade} function handles the remaining aspects of the ray tracing process. Our path tracing implementation has been verified through a pixel-wise comparison with a reference mega-kernel implementation.

\begin{figure}[tb]
    \centering
    \includegraphics[width=\linewidth]{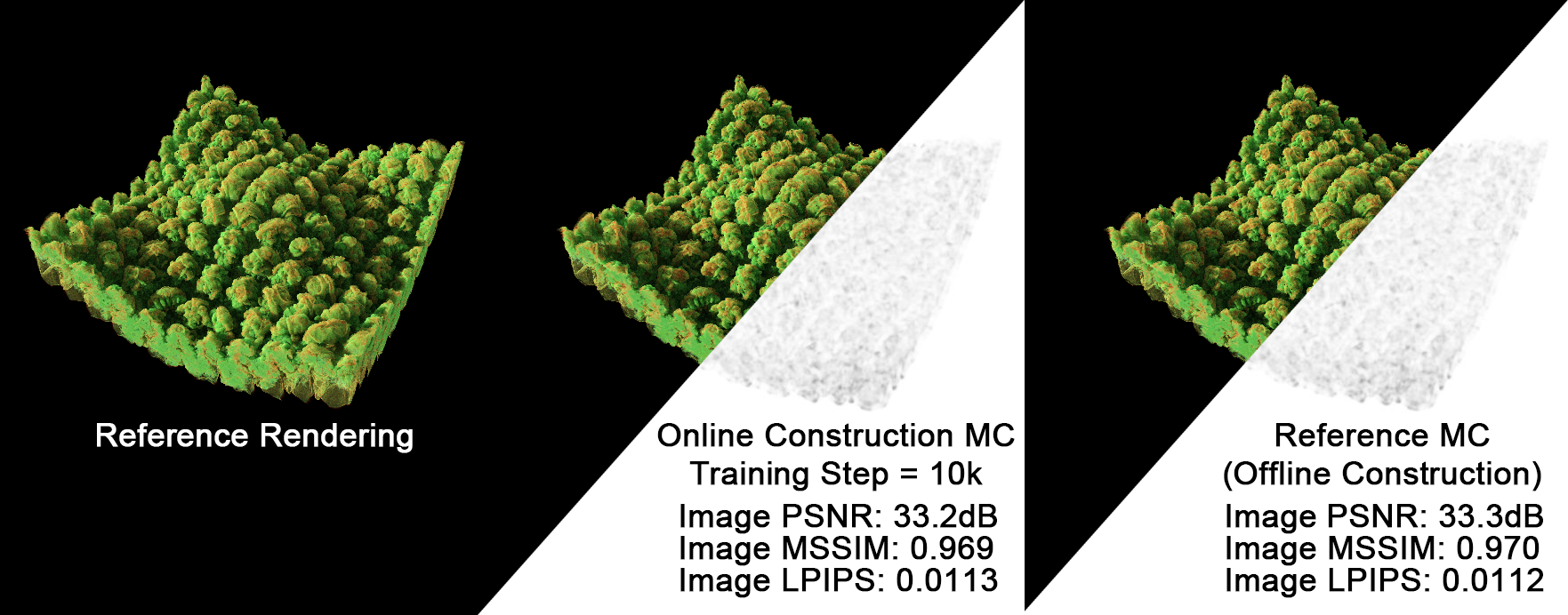}
    \vspace{-1.8em}
    \caption{Compare pre-computed and online constructed macro-cells. Left: reference rendering. Middle: rendering a neural representation (trained for 10k steps) with online constructed macro-cells. Right: rendering the same neural representation with pre-calculated macro-cells.}
    \label{vnr:fig:mc-quality}
    \vspace{-1.1em}
\end{figure}

\begin{figure*}[tb]
    \centering
    \includegraphics[width=0.9\linewidth]{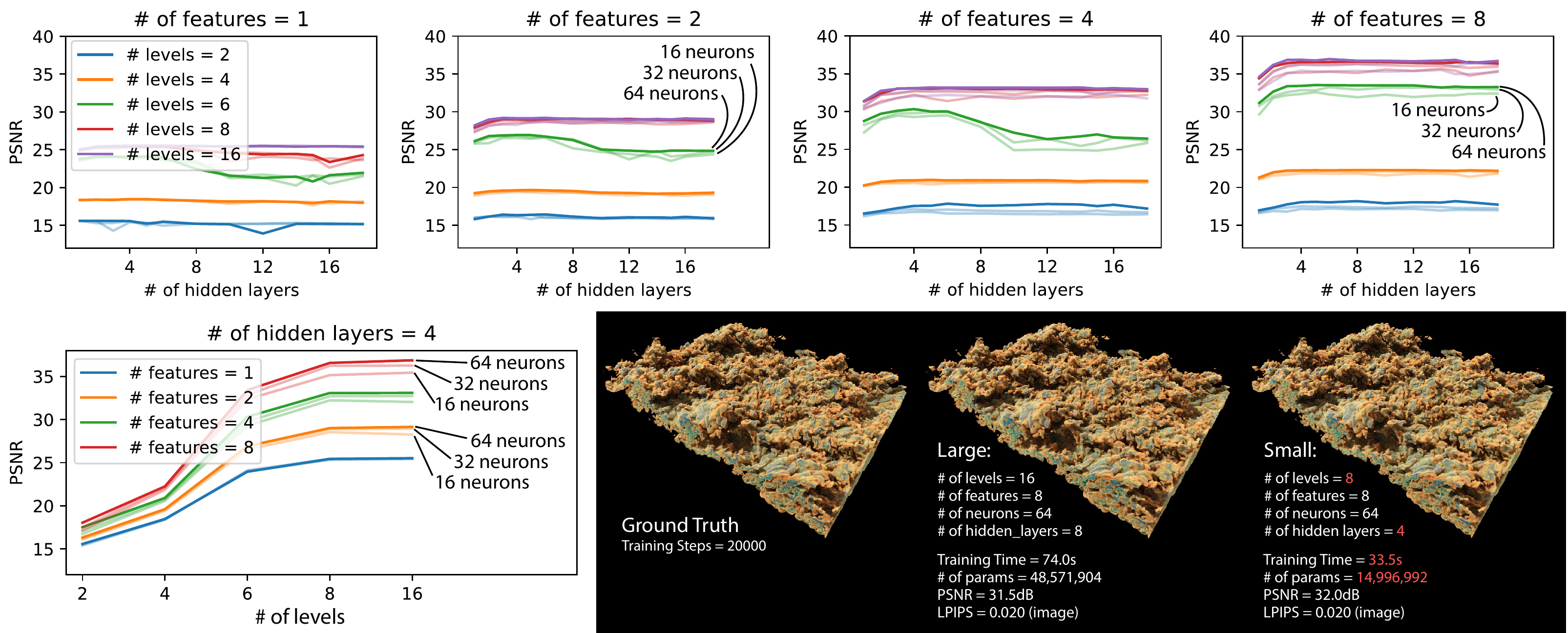}
    \vspace{-0.8em}
    \caption{Hyperparameter study. We used the same hash table size $T=2^{19}$ and scanned \# of encoding levels, \# of features per level, \# of MLP hidden layers, and \# of neurons per layer. We show all the results in the top row. We also show the result for 4-hidden-layer cases differently. Then, we picked an optimal configuration and compared it with a larger, na\"ively picked model. We compared their training times, parameter counts, volume reconstruction qualities, and rendering qualities.}
    \label{vnr:fig:paramscan}
    \vspace{-1.2em}
\end{figure*}

\subsection{Macro-Cell Optimization}

Since the cost to evaluate a volume sample using a neural representation is high, we should take as few samples as possible to maximize the performance.
To achieve that, we employ an acceleration structure called \emph{macro-cell}s~\cite{novak2014residual}.
A macro-cell contains 
${\lceil} \frac{D_x}{N_g} {\rceil}$
$\times$
${\lceil} \frac{D_y}{N_g} {\rceil}$
$\times$
${\lceil} \frac{D_z}{N_g} {\rceil}$
voxels with $D_x$, $D_y$, and $D_z$ representing the volume dimensions, and $N_g$ representing the size of a macro-cell per dimension.
In our implementation, we employ a spatial partitioning grid comprised of disjoint and closely packed macro-cells. Within each macro-cell, we store the range of values for the contained voxels and their maximum opacity $\mu_{\max}$ for the current transfer function. If any of the volume dimensions are not an integer multiple of $N_g$, our macro-cell grid may extend beyond the volume domain, which we find to be acceptable. Throughout this work, we set $N_g=64$ as we have found that this value generally strikes a favorable balance between performance and memory footprint. Although it is possible for a macro-cell to have a different $N_g$ value for each dimension, we leave the exploration of this parameter to future research. To traverse rays through macro-cells, we utilize the 3D Digital Differential Analyzer (DDA) algorithm.

The incorporation of macro-cells in path tracing can yield significant performance improvements due to the ability to construct tighter maximum opacity bounds and minimize the average difference between $\mu(t)$ and $\mu_{\max}$ employed in \Cref{alg:woodcock}. This results in a reduction in both the number of null collisions and the total number of neural network inferences. We have adopted the algorithm proposed by Hofmann~\etal~\cite{hofmann2021efficient} in our implementation and start the ray with a target optical thickness
\begin{equation}
    \tau_{\text{target}} = - log(1 - \zeta), \zeta\sim\mathcal{U}_{[0,1]}.
\end{equation}
Then we traverse the ray through the macro-cells using DDA, visiting each macro-cell in turn and accumulating its optical thickness 
\begin{equation}
    \tau_n = \sum_{i=0}^n \mu^i_{\max} \times s_i, 
\end{equation}
with $s_i$ being the length of the $i^{\text{th}}$ ray segment. 
We stop the ray when $\tau_n > \tau_{\text{target}}$ for the first time and step backward to find the exact hit.

For ray marching, we can still perform regular sampling within each cell but adaptively vary the sampling step size $\bar{s}$ based on $\mu_{\max}$. 
We use the formula recently proposed by Morrical~\etal~\cite{morrical2019efficient} to calculate 
\begin{equation}
\bar{s} = \max(s_{1} + (s_{2} - s_{1}) \cdot| \min(\mu, 1)-1 | ^p, s_{1}),
\end{equation}
where we set the minimum step size $s_{1} = 1$, maximum step size $s_{2} = 64$, and $p=2$.
Sampled opacity $\alpha$ is also corrected as $\bar{\alpha} = 1 - (1 - \alpha)^{\bar{s}/{s_{1}}}$.
When a macro-cell is transparent, we opt to bypass it entirely. It is worth noting that the utilization of adaptive sampling may cause the number of volume samples taken by each ray to fluctuate considerably. Nevertheless, this variation does not compromise the performance of our sample streaming algorithm, as we conduct a stream compaction operation to remove terminated rays after each streamout. In general, we have observed substantial speed improvements with the incorporation of macro-cells.

Typically, the value range data in macro-cells are pre-computed before rendering by iterating over all voxels.  
Such a pre-computation is also possible for online training. However, the purpose of online training would be slightly defeated, especially, when we use online training as a tool to preview large-scale data.
We propose a novel solution to this challenge. 
We observe that a neural representation is essentially a regression model constructed using all training samples. 
Thus, we can reuse the training samples generated in each training step to update the macro-cell value range field. Because the update process is embarrassingly parallelizable, the overhead is negligible.
As we do not expect to infer a neural representation in an area where no training samples have been taken, online constructed value ranges can provide sufficient accuracy for rendering.

Note that although the marco-cells are not directly constructed using values from the neural network, this will not significantly impact the rendering quality because our neural representation can already approximate the ground-truth data very accurately. Small discrepancies in macro-cells will not produce visible artifacts.
We verified this assumption by comparing rendered images with reference images and images rendered using pre-computed value ranges (\Cref{vnr:fig:mc-quality}). The model and online constructed macro-cells were trained for precisely 10k steps. Results indicated that the images rendered with online constructed macro-cells were slightly less accurate ($\sim$0.1dB in PSNR). We use online macro-cell construction in all the experiments unless stated otherwise.

%% file: contents/3.ablation.tex
\begin{figure}[tb]
    \centering
    \vspace{-1.0em}
    \includegraphics[width=0.8\linewidth]{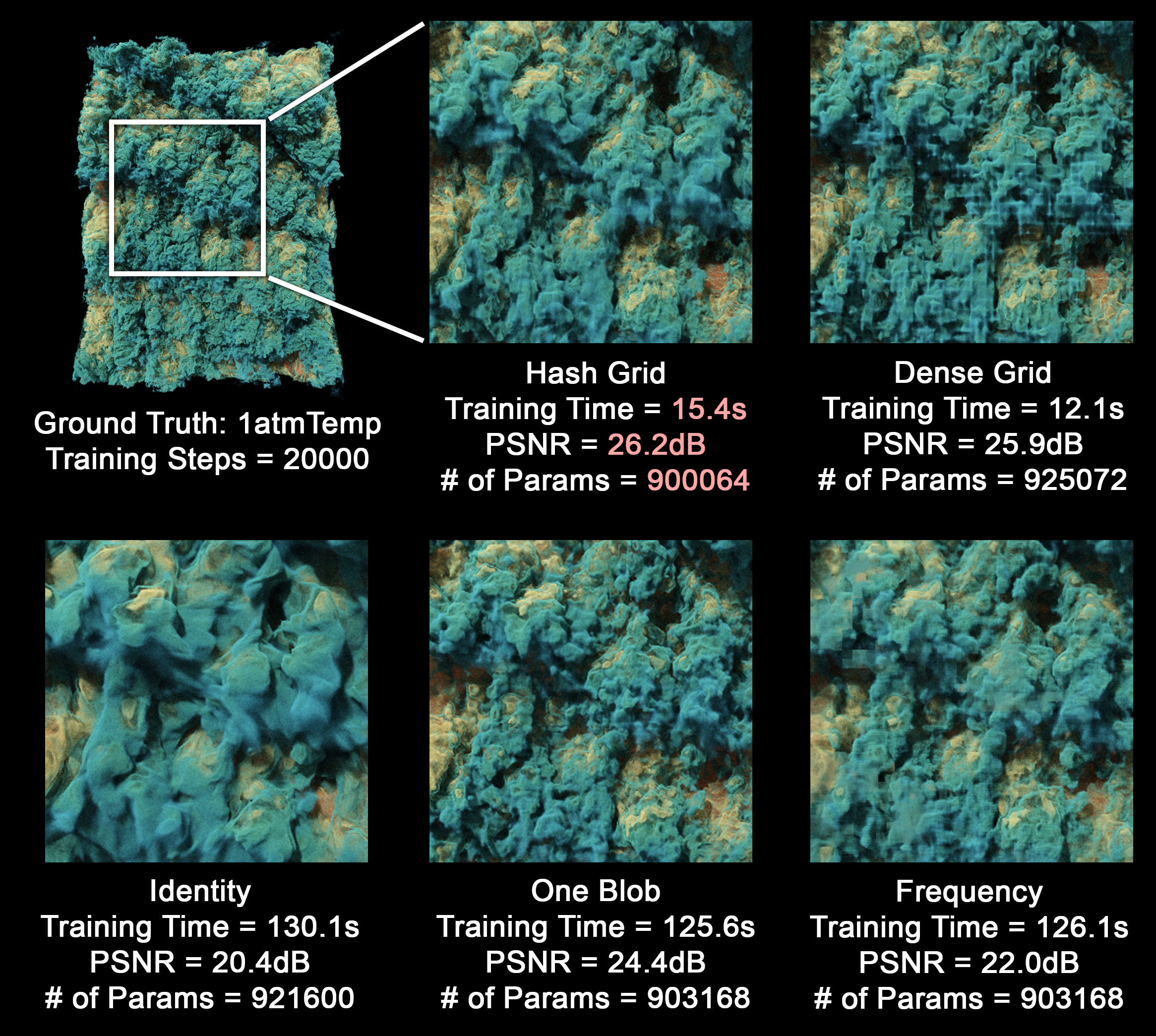}
    \vspace{-1.0em}
    \caption{Comparing similar-sized networks with different input encoding methods. For hash grid, an 8-level input encoding with 4 features per level (8$\times$4) %
    is used. For dense grid, we can only use a  2$\times$2 encoding to meet the parameter count requirement. This lead to a reduction in quality. 
    Detailed network and training configurations are specified in the appendix.
    Generally, grid-based encoding methods are faster to train. Compared with dense grid, hash grid can support more encoding layers and thus achieve better quality in equal conditions.}
    \label{vnr:fig:compare-encoding}
    \vspace{-1.0em}
\end{figure}

\section{Ablation Study}\label{vnr:sec:ablation}

It is essential to understand how the choice of encoding method and hyperparameters, such as the number of encoding features or MLP layers, would influence the model's performance. This understanding allows us to make an informed decision to choose the best combination of values.

\subsection{Input Encoding Method}

We first investigated the effectiveness of the multi-resolution hash grid encoding. To do so, we compared the hash grid encoding with four widely used encoding strategies: dense grid, one-blob~\cite{muller2019neural}, frequency~\cite{mildenhall2020nerf}, and identity (\ie no encoding). 
The dense grid encoding is similar to the hash grid encoding but uses a complete octree to store parameters.
One-blob encoding generalizes the one-hot encoding by using a Gaussian kernel to activate multiple entries.
The frequency encoding is also not trainable, and it encodes a scalar position as a sequence of sine and cosine functions.
We provide the specific network configuration for each test in \Cref{vnr:fig:compare-encoding}. 
To make the comparison fair, we used around 900,000 trainable parameters in each network. 
To match the number of parameters used by grid-based encoding methods, we used much larger MLPs for other tests. Since such a large MLP cannot fit into registers, the fast fully-fused-MLP implementation does not work. So the slower CUTLASS-based MLP implementation was used instead as a compromise. This compromise also means that the training times reported in \Cref{vnr:fig:compare-encoding} are naturally longer than other experiments performed in this work.
We used the \dataset{1atmTemp} data from a flame simulation at pressures of 1 atm created by S3D~\cite{s3d} in this study. 
We also used the heat release field (\dataset{1atmHR}) and the 10 atm variants (\dataset{10atmTemp}, \dataset{10atmHR}) from the same simulation.

The comparison between identity, frequency, and one-blob encoding indicates that mapping input coordinates to a higher-dimensional space allows the volume representation to extract more high-frequency details. This is in line with the effects observed in other domains~\cite{muller2022instant}. By comparing grid-based encoding methods with the others, we found that using a grid-based encoding could not only improve training quality, but significantly reduce training time. Finally, the comparison between two grid-based encoding methods showed that hash grid encoding could produce higher reconstruction quality and fewer stripe artifacts. This came at the cost of a slightly longer training time. However, hash grid was clearly the winner in terms of memory footprint.
Summarizing all the findings, we believe the multi-resolution hash grid encoding is the best encoding strategy for constructing volumetric neural representations.

\input{contents/table-perf.tex}

\subsection{Hyperparameter Study}\label{vnr:sec:network-params}

Next, we studied the effects of neural network parameters on network capacity.
To conduct this study, we used the \dataset{1atmHR} data and hash grid encoding with a fixed hash table size $T$. The reason to fix $T$ is that the effect of $T$ has already been well-studied by M{\"u}ller~\etal~\cite{muller2022instant}. We use the same GPU model in our work, so we refer to their findings and set $T=2^{19}$.
Then we scanned through other parameters. They are the number of encoding levels, the number  of  features per encoding level, the number of neurons per MLP layer, and the number of hidden layers in MLP. We report the reconstruction quality after training in terms of PSNR.
For each configuration, the network was optimized for a sufficient amount of steps (at least 200k) until it practically converges.
We believe the PSNR computed at this point reflects the network capacity.
\Cref{vnr:fig:paramscan} illustrates our results.

From the line plots shown in \Cref{vnr:fig:paramscan}, we can see that two encoding parameters (\# of levels, \# of features) produced a more significant impact on network capacity. This is understandable as most of the trainable parameters are stored in the encoding layer. However, both parameters also suffered from the law of diminishing returns. For the tested data, having 4-8 features per level and 4-8 encoding levels seems to be the sweet spot. Beyond that, the extra benefits were limited. As for the MLP, it seems that having 4-8 hidden layers was generally enough. Having more hidden layers did not bring any extra performance benefits and could even lead to performance reductions. Having more neurons in each MLP layer was generally beneficial, but the benefits were limited. However, this was likely because the MLPs used were all relatively small and thus only produced a limited impact.

Additionally, smartly picking network parameters could sometimes significantly improve training time and memory footprint for our volumetric neural representation.
In \Cref{vnr:fig:paramscan}, we compared two networks, optimized under identical conditions, for their reconstruction qualities using PSNR of reconstructed volumes, perceptual qualities using LPIPS~\cite{zhang2018perceptual} of rendered images, training time, and memory footprint in terms of parameter count.
We constructed the \emph{large} network with the maximum number of encoding levels, the maximum number of features per encoding level, the maximum number of neurons per MLP layer, and a high number of hidden layers. The \emph{small} network reduced the number of encoding levels and MLP hidden layers by 50\% following our previous findings. 
Our \emph{small} network achieved nearly the same quality. However, it required less than half of the training time and less than a quarter of the memory footprint.

%% file: contents/table-perf.tex
\begin{table*}[tb]
\caption{Comparison of training results between our method and two related techniques: \fvsrn, which was adjusted to match our models' compression ratios, and \tthresh, which was adjusted to match our models' PSNRs after 20k steps of training. Our models were trained on an NVIDIA RTX 3090. \fvsrn models were trained on an NVIDIA RTX 3090Ti. \tthresh experiments were run on an 88-core server with 256GB RAM as it is a CPU-based algorithm. OOM stands for out of memory. We highlight best results within each category with \best{underline}.}\label{vnr:tab:quality}
\scriptsize\centering
\vspace{-1em}
\begin{tabu}{rc|ccc|ccc|cc|cc|ccc}\toprule
&&\multicolumn{3}{c|}{PSNR $\uparrow$} 
& \multicolumn{3}{c|}{MSSIM $\uparrow$} 
& \multicolumn{2}{c|}{Size (\textit{MB}) $\downarrow$} 
& \multicolumn{2}{c|}{Ratio $\downarrow$} 
& \multicolumn{3}{c }{Time (\textit{s}) $\downarrow$} \\
&&\multicolumn{2}{c}{Ours}&\fvsrn&\multicolumn{2}{c}{Ours}&\fvsrn&Ours&\tthresh&Ours&\tthresh&Ours&\tthresh&\fvsrn\\
Dataset       &Dimensions           & 200k      & 20k&           & 200k       &20k  &            &           &           &                 &                  &           &           &          \\\midrule
{RM-T60}      &   (256, 256, 256)   &\best{36.5}&34.5&\last{25.4}&\best{0.991}&0.986&\last{0.947}&\best{1.0} & 1.6       &\best{67$\times$}& 41$\times$       &7.0        &\best{3.4} &\last{100}\\
{Skull}       &   (256, 256, 256)   &\best{41.9}&40.6&\last{34.7}&\best{0.987}&0.984&\last{0.961}&\best{1.0} & 2.0       &\best{67$\times$}& 34$\times$       &6.9        &\best{3.6} &\last{94} \\
{1atmTemp}    &  (1152, 320, 853)   &\best{38.4}&34.4&\last{30.1}&\best{0.981}&0.964&\last{0.944}&  44.6     &\best{9.2} & 28$\times$      &\best{136$\times$}&\best{39.6}&126        &\last{418}\\
{10atmTemp}   &  (1152, 426, 853)   &\best{36.4}&32.4&\last{28.1}&\best{0.982}&0.965&\last{0.939}&  44.6     &\best{9.5} & 38$\times$      &\best{176$\times$}&\best{39.5}&259        &\last{310}\\
{1atmHR}      &  (1152, 320, 853)   &\best{36.8}&31.9&\last{26.5}&\best{0.992}&0.976&\last{0.953}&  28.6     &\best{9.5} & 44$\times$      &\best{133$\times$}&\best{31.7}&127        &\last{227}\\
{10atmHR}     &  (1152, 426, 853)   &\best{32.9}&28.1&\last{23.9}&\best{0.983}&0.957&\last{0.924}&  28.6     &\best{20.5}& 59$\times$      &\best{82$\times$} &\best{31.6}&\last{260} &227       \\
{Chameleon}   &  (1024, 1024, 1080) &\best{54.3}&49.8&\last{43.6}&\best{0.998}&0.997&\last{0.991}&  28.6     &\best{7.7} & 158$\times$     &\best{587$\times$}&\best{24.8}&\last{789} &194       \\
{MechHand}    &   (640, 220, 229)   &\best{41.5}&39.1&\last{27.0}&\best{0.998}&0.997&\last{0.967}&  6.7      &\best{3.8} & 19$\times$      &\best{34$\times$} &5.8        &\best{12.9}&\last{178}\\
{SuperNova}   &   (432, 432, 432)   &\best{52.2}&49.9&\last{41.7}&\best{0.998}&0.997&\last{0.988}&  8.6      &\best{4.1} & 37$\times$      &\best{79$\times$} &\best{17.9}&64.6       &\last{120}\\
{Instability} & (2048, 2048, 1920)  &\best{31.6}&27.9&\last{OOM} &\best{0.961}&0.941&\last{OOM}  &\best{52.6}& OOM       &612$\times$      & OOM              &\best{96.5}&\last{OOM} &\last{OOM}\\
{PigHeart}    & (2048, 2048, 2612)  &\best{55.0}&52.9&\last{OOM} &\best{0.997}&0.996&\last{OOM}  &\best{6.6} & OOM       &6640$\times$     & OOM              &\best{78.5}&\last{OOM} &\last{OOM}\\
{DNS}         & (10240, 7680, 1536) &\best{35.9}&34.6&\last{OOM} &\best{0.922}&0.917&\last{OOM}  &\best{1520}& OOM       &320$\times$      & OOM              &\best{6706}&\last{OOM} &\last{OOM}\\\midrule
{PH-LDR}      & (2048, 2048, 2612)  &\best{41.4}&40.1&\last{OOM} &\best{0.938}&0.931&\last{OOM}  &\best{38.6}& OOM       &1135$\times$     & OOM              &\best{88.4}&\last{OOM} &\last{OOM}\\\bottomrule
\end{tabu}
\vspace{-1em}
\end{table*}

\begin{table}[tb]
\vspace{-0.5em}
\caption{Additional training results. Our fix-sized networks were trained for 20k steps. \neurcomp networks were trained for 75 epochs. Online training experiments were performed using an NVIDIA RTX 3090.}
\vspace{-1em}
\label{vnr:tab:training-extra}
\scriptsize\centering%
\begin{tabu}{r|rcc|crr}
\toprule 
& \multicolumn{3}{c|}{Fix-Sized Network} & \multicolumn{3}{c}{Online Training} \\
& \multicolumn{2}{c}{\makecell{Ours\\(29.1MB)}} & \makecell{\neurcomp\\(\best{3.6MB})} 
& \multicolumn{3}{c}{(29.1MB) ~ Latency} \\
\cmidrule(lr){2-3} \cmidrule(lr){4-4} \cmidrule(lr){5-7}
Dataset       &Time            & PSNR        & PSNR      & FPS & Train & Render \\\midrule
{1atmTemp   } & 31.2\textit{s} & \best{34.6} & 33.4      & 14  & 2.0\textit{ms} &  68\textit{ms}  \\
{10atmTemp  } & 31.0\textit{s} & \best{32.3} & 30.1      & 17  & 2.0\textit{ms} &  58\textit{ms}  \\
{1atmHR     } & 31.6\textit{s} & \best{32.0} & 28.8      & 13  & 1.7\textit{ms} &  74\textit{ms}  \\
{10atmHR    } & 31.7\textit{s} & \best{28.2} & 27.0      & 8.0 & 1.6\textit{ms} & 125\textit{ms}  \\
{Chameleon  } & 25.6\textit{s} & 49.7        &\best{50.1}& 42  & 1.4\textit{ms} &  22\textit{ms}  \\
{MechHand   } & 26.6\textit{s} & 41.4        &\best{46.2}& 16  & 1.0\textit{ms} &  60\textit{ms}  \\
{SuperNova  } & 25.8\textit{s} & \best{51.4} & 45.3      & 33  & 1.0\textit{ms} &  29\textit{ms}  \\
{Instability} & 83.2\textit{s} & \best{27.9} &  OOM      & 10  & 5.0\textit{ms} &  95\textit{ms}  \\
{PigHeart   } & 84.0\textit{s} & \best{54.3} &  OOM      & 17  & 3.8\textit{ms} &  56\textit{ms}  \\
{DNS        } & 5940\textit{s} & \best{32.8} &  OOM      & 2.7 & 281\textit{ms} &  98\textit{ms}  \\
\bottomrule
\end{tabu}%
\vspace{-1em}
\end{table}

%% file: contents/6.evaluation.tex
\begin{figure*}[tb]
    \centering
    \includegraphics[width=\linewidth]{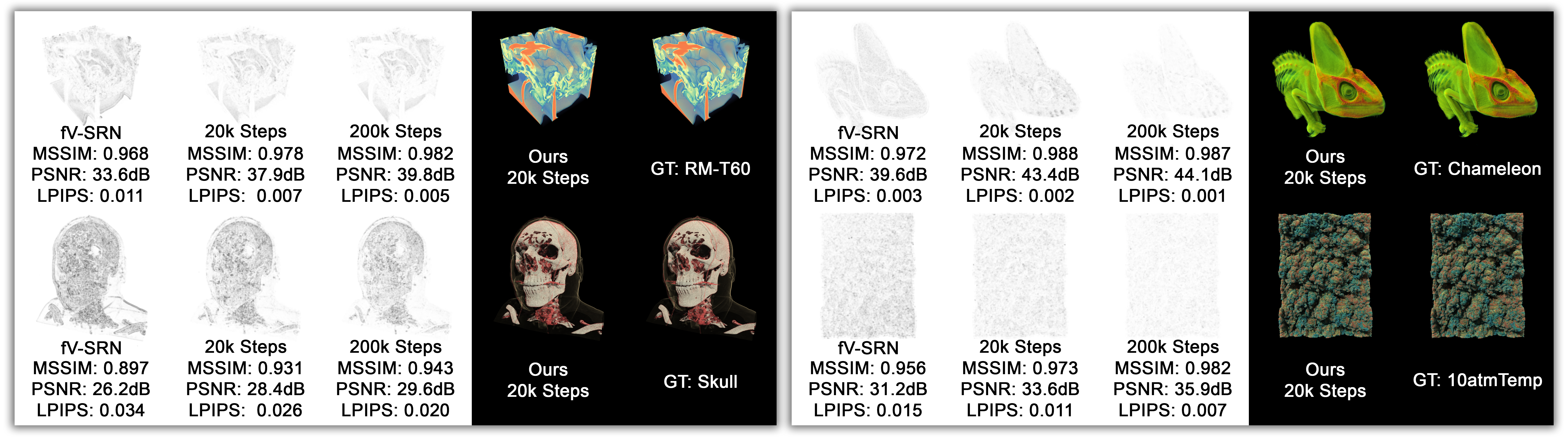}
    \vspace{-2.5em}
    \caption{Comparisons of rendering quality between \fvsrn and our models. Neural representations obtained from the previous experiment shown in~\Cref{vnr:tab:quality} were used. We rendered each neural representation by exactly 200 frames. The image space PSNR, SSIM, and LPIPS~\cite{zhang2018perceptual} were computed against reference renderings. \dataset{RM-T60} models were rendered using regular ray marching. \dataset{Skull} models were rendered using ray marching with single-shot shadows. \dataset{Chameleon} and \dataset{10atmTemp} models were rendered using path tracing with a single light source.}
    \label{vnr:fig:renderings}
    \vspace{-1em}
\end{figure*}

\section{Performance Evaluation}

We evaluated our implementations on a Windows machine with an NVIDIA RTX 3090, an Intel 10500, 64GB RAM, and NVMe Gen3 SSDs.
We used several datasets with various sources, contents, and sizes. Because \dataset{Instability}~\cite{osv} and \dataset{PigHeart}~\cite{osv} could not be loaded into GPU memory, we used the OpenVKL-based variant to generate samples. Our out-of-core sampling technique was used for the 0.95TB channel flow \dataset{DNS} dataset~\cite{dns}.

\subsection{Training}\label{vnr:sec:training}

We evaluated the training process by investigating the training time, model size, and volume reconstruction quality (PSNR and SSIM). A small hyperparameter scan was performed on each dataset to find a good neural representation.
We scanned MLPs with 16, 32, and 64 neurons with 2-6 hidden layers.
We also scanned the hash grid encoding with 4 and 8 features per level, the number of levels from 6-12, and the hash table size from $2^{16}$ to $2^{19}$ by incrementally changing the power term.
We trained each configuration for 2000 steps (equivalent to 3-10s) and selected the best-performing configuration in terms of PSNR.
In practice, performing such a parameter scan is not strictly necessary, as a guess based on our ablation study (\Cref{vnr:sec:ablation}) is often good enough. 
To verify this, we also trained all the datasets using a single configuration (detailed in the appendix) and recorded their performance and training time. We discuss the results in \Cref{vnr:sec:reconstruct-quality}.

After identifying appropriate settings, we optimized two sets of neural representations on each dataset. 
The first set was optimized for 200k steps. All the models were extensively optimized at this point.
Then, we created the second set of models with the same configuration but optimized them with 10$\times$ fewer training steps (20k-steps). 
This was to study how well the models could perform with a limited training budget.
The training times of all the 20k-step models are reported in \Cref{vnr:tab:quality}.
We also report each volume's model size and the calculated compression ratio.

Although our out-of-core sampling method enables us to train the terascale \dataset{DNS} dataset efficiently, it is still significantly slower than other in-core training methods. Using the brute force method to find the best configuration would be too costly. Therefore, we manually identified a good configuration for \dataset{DNS}. We also used L2 loss because we found it to provide more training stability in conjunction with an increased number of training parameters.

\subsubsection{Reconstruction and Rendering Quality}\label{vnr:sec:reconstruct-quality}

We measured how well our volumetric neural representation could learn features from a given volume data by comparing the reconstructed volume (\Cref{vnr:tab:quality}) and rendered images (\Cref{vnr:fig:renderings}) with the ground truth.
The appendix can find rendering results for datasets not shown in \Cref{vnr:fig:renderings}. 

We progressively decoded the volume at its original resolution and calculated PSNR and the mean structural similarity index (MSSIM) between the reconstructed volume and the original data.
Results for both 20k and 200k models are reported in \Cref{vnr:tab:quality}.
We also rendered both models using our sample streaming path tracer for 200 frames and compared image differences in \Cref{vnr:fig:renderings} in terms of PSNR, MSSIM, and LPIPS.
Our neural representations (200k step models) could provide good volume reconstruction and rendering accuracy. With a limited training budget (20k step models), our neural representations could still perform reasonably well, especially for tasks like rendering, where decreases in image accuracy were tiny.

As mentioned previously, we also trained each dataset with a fixed network configuration. Results are shown in \Cref{vnr:tab:training-extra}. We could still get nearly identical reconstruction quality without a brute-force parameter scan. The training times were slightly different compared to results in \Cref{vnr:tab:quality}. This is expected as the number of parameters changed significantly for some data.

\add{In our study of the \dataset{DNS} dataset, we observed that only about 10\% of the total voxels were utilized as training samples, despite 200k training steps. Interestingly, this limited sample size did not hinder the neural network's performance; it demonstrated high reconstruction quality, reflected in a PSNR of approximately 35dB. These results suggest our neural representation possesses commendable predictive accuracy for unseen coordinates. This outcome can potentially be attributed to two factors. Firstly, the adoption of multiple encoding resolutions allows our neural network to independently model features at various scales, potentially improving data reconstruction quality. Secondly, the inherent properties of the dataset may facilitate easier approximation of its features. Nevertheless, a comprehensive evaluation of this capability extends beyond the scope of this paper, hence we earmark it for future exploration.}

\begin{figure}[tb]
\centering
\includegraphics[width=0.9\linewidth]{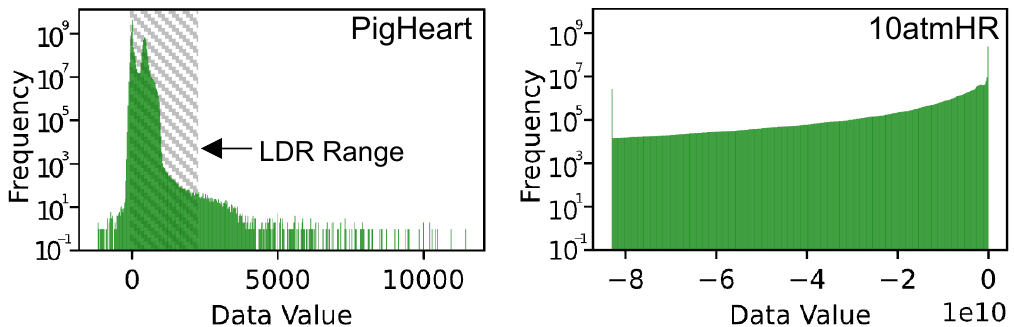}
\vspace{-1.2em}
\caption{The value distribution of \dataset{PigHeart} (left), the low-dynamic-range version (LDR) of \dataset{PigHeart} (left, shaded region), and \dataset{10atmHR} (right).}
\label{vnr:fig:data-histogram}
\end{figure}

\subsubsection{Comparison with \fvsrn}

We conducted a comprehensive comparison between our method and the \fvsrn method~\cite{weiss2021fast} with \add{latent grid}. To ensure a fair assessment, we trained \fvsrn networks on a Linux machine equipped with an NVIDIA RTX 3090Ti using the same number of trainable parameters as our 20k/200k models. 
\add{Additionally, to make our comparison more relavent, we have incorporated all publicly accessible datasets (\ie \dataset{RM-T60} and \dataset{Skull}) that were utilized in the study conducted by Weiss~\etal~\cite{weiss2021fast}. We adhered to the recommended \fvsrn configuration for these datasets. Despite their relatively small scale, these two datasets remain interesting because they share the same network configuration with the other datasets presented in Weiss~\etal's report.}

We evaluated the effectiveness of the two methods by assessing various factors, including training times, reconstruction quality, and rendering quality, as presented in both~\Cref{vnr:tab:quality} and~\Cref{vnr:fig:renderings}. 
\add{Compared with \fvsrn, our models were significantly faster to train.}
Notably, \add{\fvsrn's customized CUDA kernel is designed specifically for accelerating network inference performance, while its training system is still implemented using PyTorch,} which may have further exaggerated the training time differences. 
Despite this, our models consistently outperformed the \fvsrn approach in all \add{the quality} metrics, even when utilizing the same number of parameters.
This result may be attributed to our improved network design. The use of multi-resolution hash-grid encoding can not only accelerate the overall training time but also reduce the number of training steps required to achieve the desired quality.
It is noteworthy that the \fvsrn training system generated $256^3$ samples per epoch, and each model was optimized for 200 epochs, equivalent to training our models for 51.2k steps.

\begin{figure}[tb]
    \centering
    \includegraphics[width=0.9\linewidth]{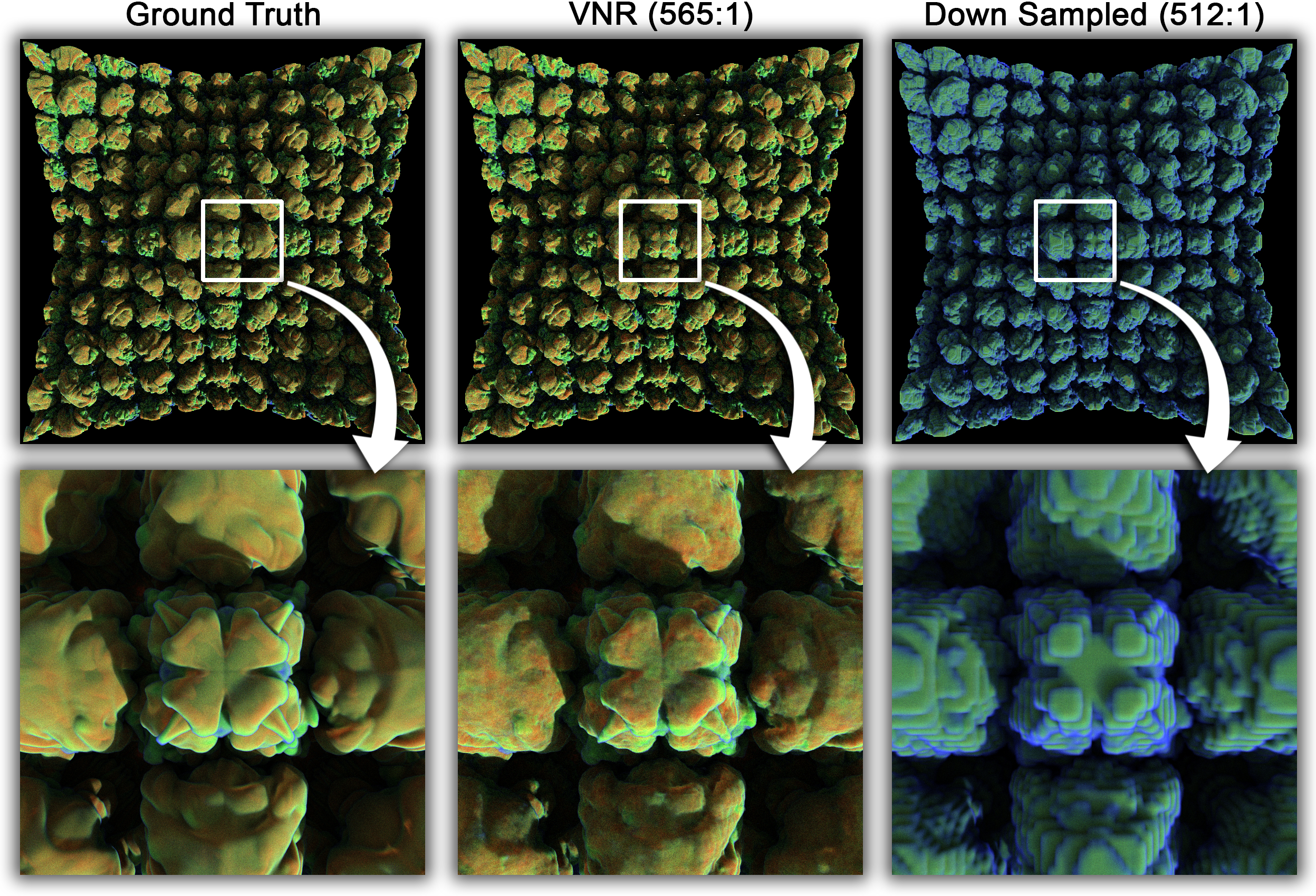}
    \vspace{-1.2em}
    \caption{Rendering quality comparison between ground truth data and a trained volumetric neural representation (10k training steps, same transfer function). 
    Left: the ground-truth rendering (generated on a different GPU with more VRAM).
    Middle: path tracing of the neural representation (VNR). Artifacts are relatively small. 
    Right: the rendering of the 8$\times$ down-sampled volume (to match the VNR compression ratio).
    The down-sampled volume produced very significant artifacts.}
    \label{vnr:fig:downsample-vs-inshader}
    \vspace{-1em}
\end{figure}

\subsubsection{Compression Ratio}
In \Cref{vnr:tab:quality}, 
we highlight the memory footprints of our neural representations and the corresponding compression ratios. 
The \dataset{DNS} data is double precision, whereas others are single-precision floating-point volumes. 
Overall, our method could faithfully compress data by 10-1000$\times$, especially for large datasets such as \dataset{Chameleon} (4.5GB), \dataset{Instability} (32GB), \dataset{PigHeart} (44GB), and \dataset{DNS} (950GB).

Notably, the neural representation optimized for the \dataset{PigHeart} data presented a surprisingly high compression ratio (6640:1). Our investigation suggests that the \dataset{PigHeart} data has a fairly uneven value distribution (left image in \Cref{vnr:fig:data-histogram}). This will make the reconstruction error relatively minor compared with more evenly distributed data such as \dataset{10atmHR} (right image in \Cref{vnr:fig:data-histogram}).
To verify our claim, we clamped the dynamic range of \dataset{PigHeart} volume to $[0, 2000]$ (referred to as \dataset{PH-LDR}, with its data distribution highlighted as the shaded area in \Cref{vnr:fig:data-histogram}). Then we trained a neural representation with the same configuration. We discovered that the low-dynamic-range version produced a lower reconstruction quality.

\subsubsection{Comparison with \tthresh and \neurcomp}

We also compared our method with \neurcomp~\cite{lu2021compressive} and the state-of-the-art volume compression technique \tthresh~\cite{ballester2019tthresh}.

For \tthresh, each dataset was compressed to the same PSNR as our 20k-step model.
As \tthresh is a CPU-based algorithm, we performed all the compressions on a dual Intel Xeon E5-2699 (88-core) server with 256GB main memory. We report our results in \Cref{vnr:tab:quality}.
We found that \tthresh performed much better in terms of compression ratio. However, it took the algorithm significantly longer to compress most datasets, and the compressed volume cannot be accessed or rendered without an explicit decompression.

For \neurcomp, all datasets were trained using eight residual blocks, each comprising 256 neurons, on an NVIDIA RTX 3090. Despite our best efforts, it was challenging to balance the number of parameters between \neurcomp and our methods, as optimizing larger \neurcomp networks with 300-1000 neurons per block proved to be exceedingly difficult. This can partly be attributed to the lack of input encoding in \neurcomp. Nevertheless, our experiments have shown that a 256$\times$8 \neurcomp network could achieve comparable reconstruction quality \add{compared with} our fixed-sized network, while delivering significantly better compression ratios, as demonstrated in \Cref{vnr:tab:training-extra}. However, all \neurcomp networks required multiple hours of training on any given dataset. Even after accounting for the performance differences between Python and CUDA, the discrepancies in training time remained significant. As such, we have opted to exclude the training times for \neurcomp from our comparison.

\subsubsection{Training Time} 

We compared the training of 20k models shown in \Cref{vnr:tab:quality}. In general, we observed two trends.
Firstly, smaller models were generally faster to train. This is because there are fewer gradients to compute in each training step.
Secondly, \dataset{Instability}, \dataset{PigHeart}, and \dataset{DNS} data were significantly slower to train. This is because they are trained with CPU-based sampling and out-of-core sampling methods.
For CPU-based sampling, the sampling process was slower, and the overhead of constantly copying training samples from CPU to GPU was not negligible.
For out-of-core sampling, the performance was primarily bounded by the disk I/O latency.
We further investigated this and compared the out-of-core sampling method with the method directly using virtual memory via the \texttt{CreateFileMapping} API (with uniformly sampled random coordinates). 
We used both methods with trilinear interpolation to train the same model for 10k steps. The virtual memory method took 12.7 hours to complete. Our out-of-core sampling method only took 54 mins ($\sim$14$\times$ faster).
This is because randomly and uniformly accessing coordinates would lead to a considerable number of page faults, which in turn cause I/O overhead.

\begin{figure}[tb]
    \centering
    \includegraphics[width=\linewidth]{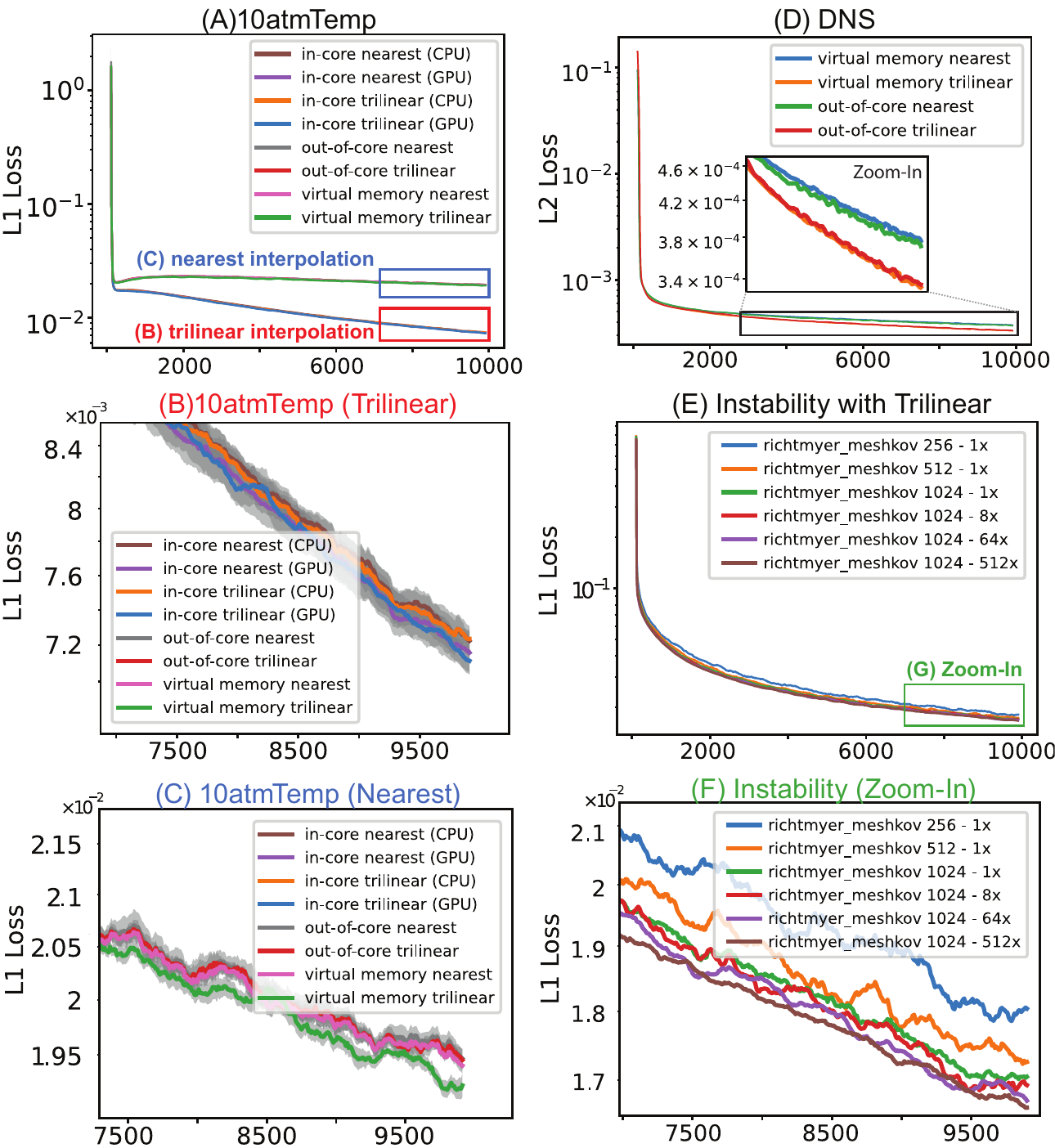}
    \vspace{-2em}
    \caption{(A-C) The loss curves of the \dataset{10atmTemp} using different sampling methods. Each dataset was trained for 10k steps.
    (D) \dataset{DNS} data using different out-of-core sampling methods. (E-F) \dataset{Instability} data using the out-of-core sampling method but with different parameters. For each line, the first number indicates the number of blocks to refresh $S$; the second number indicates $\frac{R}{S}$, where $R$ is the total number of blocks stored in the system memory.}
    \label{vnr:fig:training-methods}
    \vspace{-1em}
\end{figure}

In \Cref{vnr:tab:training-extra}, we also demonstrate the performance of online training. As we can see, for all the datasets, interactive performance was achieved. 
The breakdown of training and rendering latency indicates that, with our settings, all the online training processes were rendering-bound, with each training step only making up for a fraction of the total time.
The only exception was \dataset{DNS}, which used the out-of-core sampling method for training. However, this is understandable as this technique requires a lot of high-latency I/O operations.

In \Cref{vnr:fig:downsample-vs-inshader}, we compared a neural representation (optimized for 10k steps, configuration in the appendix) with a na\"ively downsampled version of the same dataset. The downsampling factor was tuned to achieve a similar level of compression. We compared the rendering of both versions to the ground truth. Renderings were done using our sample streaming path tracer.
The image rendered from our neural representation could generally match the ground truth result, with minor artifacts visible only after zooming in.
In contrast, the downsampled volume failed to capture many details of the data and produced a noticeable color shift. This was because high- and low-value features could be averaged out during downsampling.

\subsubsection{Sampling Pattern}\label{vnr:eval:sampling} 

Different sampling strategies might lead to differences in training quality. To study this relationship, we conducted three experiments.
In the first one, we trained the \dataset{10atmTemp} data with the exact neural representation using in-core, out-of-core (default settings as mentioned in \Cref{vnr:sec:training-ofc}), and virtual memory sampling methods. For each sampling method, we tested both trilinear and nearest neighbor interpolations. The loss curve of each training could be found in \Cref{vnr:fig:training-methods}A-C. We observed that experiments using the same interpolation method all generally matched each other.  However, trilinear interpolation allowed training to perform significantly better than their nearest neighbor interpolation counterparts.
Next, we investigated the large \dataset{DNS} dataset. Due to its size, we performed only out-of-core (default settings) and virtual memory sampling methods. Results are shown in \Cref{vnr:fig:training-methods}D. Although the out-of-core sampling method did not generate sample coordinates uniformly and randomly, it did not obviously impact the training.
Finally, we trained the \dataset{Instability} with different out-of-core training settings ($M$ and $R$). 
We found that larger values for $M$ and $R$ both positively impacted training quality.

\begin{table*}[tb]
\caption{Comparison of framerates between \fvsrn and our method for three rendering methods. \textit{MC} stands for macro-cell, \textit{IS} for in-shader inference, and \textit{SS} for sample streaming. Note that some \fvsrn trainings ran out of memory, thus marked as OOM. Fastest framerates within each category are highlighted with \best{underline}. All models were obtained from the experiment described in \Cref{vnr:tab:quality} and rendered using an NVIDIA RTX 3090.}
\vspace{-1em}
\label{vnr:tab:volume-fps}
\scriptsize\centering
\begin{tabu}{r|ccc cc|ccc cc|ccc cc}
\toprule
& \multicolumn{5}{c|}{Ray Marching}
& \multicolumn{5}{c|}{Ray Marching w/ Single Shot Shadows}
& \multicolumn{5}{c }{Path Tracing}\\
\cmidrule(lr){2-6} \cmidrule(lr){7-11} \cmidrule(lr){12-16} 
& \multicolumn{3}{c }{w/o MC}
& \multicolumn{2}{c|}{w/  MC} 
& \multicolumn{3}{c }{w/o MC}
& \multicolumn{2}{c|}{w/  MC} 
& \multicolumn{3}{c }{w/o MC}
& \multicolumn{2}{c }{w/  MC} \\
\cmidrule(lr){2-4} \cmidrule(lr){5-6} \cmidrule(lr){7-9} \cmidrule(lr){10-11} \cmidrule(lr){12-14} \cmidrule(lr){15-16}     
Dataset
& \fvsrn & IS & SS & IS & SS
& \fvsrn & IS & SS & IS & SS
& \fvsrn & IS & SS & IS & SS \\\midrule
RM-T60      &\last{38 } & 89       & 133 & 108 & \best{204} & \last{30}  & 69       & 102 & 86  &\best{151}&\last{4.5 } & 13        & 12   &\best{30}& 28       \\
Skull       &\last{9.5} & 19       & 35  & 128 & \best{158} & \last{7.8} & 16       & 26  & 98  &\best{123}&\last{1.7 } & 4.6       & 5.8  &\best{16}& 16       \\
1atmTemp    & 5.6       &\last{4.6}& 13  & 20  & \best{69}  & 4.1        &\last{3.4}& 7.4 & 13  &\best{42} & 0.52       &\last{0.37}& 2.9  & 2.0     &\best{14 }\\
10atmTemp   & 5.9       &\last{4.8}& 14  & 18  & \best{60}  & 4.2        &\last{3.5}& 7.9 & 12  &\best{37} & 0.58       &\last{0.37}& 2.8  & 2.5     &\best{17 }\\
1atmHR      &\last{5.7} & 8.4      & 18  & 30  & \best{89}  & \last{4.1} & 6.0      & 9.5 & 20  &\best{55} &\last{0.54} & 0.58      & 2.9  & 3.2     &\best{14 }\\
10atmHR     &\last{4.6} & 6.5      & 12  & 24  & \best{67}  & \last{3.5} & 4.8      & 7.6 & 14  &\best{43} &\last{0.49} & 0.53      & 2.9  & 2.0     &\best{8.2}\\
Chameleon   &\last{0.8} & 1.2      & 2.0 & 10  & \best{18}  & \last{0.6} & 1.1      & 1.6 & 8.1 &\best{15} &\last{0.19} & 0.25      & 0.79 & 14      &\best{42 }\\
MechHand    &\last{4.5} & 6.5      & 12  & 19  & \best{75}  & \last{3.1} & 5.0      & 9.6 & 15  &\best{54} &\last{0.91} & 1.3       & 3.4  & 5.3     &\best{10 }\\
SuperNova   &\last{2.2} & 2.9      & 4.4 & 24  & \best{35}  & \last{1.7} & 2.6      & 3.9 & 19  &\best{29} &\last{0.63} & 0.83      & 2.0  & 11      &\best{32 }\\
Instability &\last{OOM} & 0.57     & 1.7 & 24  & \best{79}  & \last{OOM} & 0.49     & 1.2 & 18  &\best{53} &\last{OOM}  & 0.09      & 0.43 & 2.8     &\best{10 }\\
PigHeart    &\last{OOM} & 0.80     & 2.4 & 30  & \best{82}  & \last{OOM} & 0.70     & 1.8 & 22  &\best{57} &\last{OOM}  & 0.16      & 0.77 & 4.9     &\best{20 }\\
DNS (dp)    &\last{OOM} & 0.58     & 1.4 & 17  & \best{39}  & \last{OOM} & 0.49     & 1.1 & 12  &\best{25} &\last{OOM}  & 0.11      & 0.63 & 2.0     &\best{11 }\\
\bottomrule
\end{tabu}
\vspace{-0.6em}
\end{table*}

\subsection{Rendering}

In \Cref{vnr:tab:volume-fps}, we report the rendering performance of both rendering techniques we developed---in-shader and sample streaming. For each technique, three rendering modes were visited---ray marching with the emission-absorption lighting model (in the appendix), ray marching with single-shot heuristic shadows, and volumetric path tracing with next-event estimation. 
We recorded the performance with and without macro-cell optimization for each rendering mode.
\add{The frame buffer size was set at 768$\times$768. Prior to rendering, all datasets were scaled so that each voxel would correspond to a unit cube in the world coordinate system. For ray marching methods, we set the sampling step size to 1. For path tracing methods, we used a Russian roulette depth of 4.}
We also used the same %
camera angle and transfer function for each dataset for experiments in \Cref{vnr:fig:renderings} and \Cref{vnr:tab:training-extra}. 

From the results, we derived two main insights. Firstly, sample streaming methods were 3-8$\times$ faster than in-shader rendering methods because the sample streaming method could allow the network inference step to exploit more parallelism using extra GPU threadblocks. However, this is generally impossible for in-shader rendering, as one ray is managed by exactly one GPU thread.
Secondly, macro-cells also brought significant speedups in performance. For in-shader methods, we observed 2-10$\times$ speedups. For sample streaming methods, we could get speedups as high as 40$\times$. This is because macro-cells allow the renderer to traverse through empty and low-opacity regions quickly; thus, fewer volume samples are queried. For rendering neural representations, this is crucial as the cost to compute each volume sample is very high.

\begin{figure}[tb]
    \centering
    \includegraphics[width=0.9\linewidth]{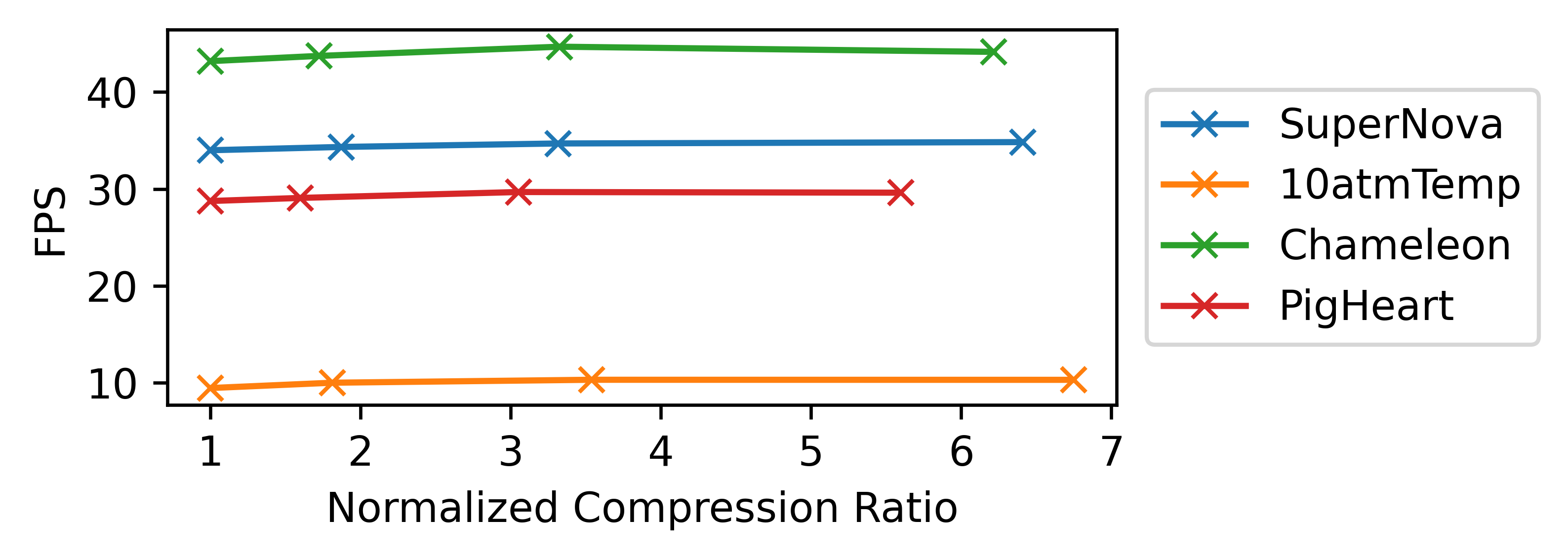}
    \vspace{-1.2em}
    \caption{\add{The correlation between compression ratio and rendering speed. For each dataset, we normalized all compression ratios with respect to the corresponding compression ratio reported in \Cref{vnr:tab:quality}. This experiment was performed using an NVIDIA RTX 8000 GPU.}}
    \vspace{-1em}
    \label{vnr:fig:fps-scaling}
\end{figure}

\add{In addition, we examined the relationship between compression ratio and rendering speed. We conducted this investigation utilizing models presented in \Cref{vnr:tab:quality} (referred to as base models), and then reduced the hash table size $T$ of each model exponentially via the \texttt{log2\_hashmap\_size} parameter. The act of reducing $T$ efficiently decreases model sizes, given that the majority of the trainable parameters in our models are stored within hash tables. In contrast, alterations to the MLP size would not significantly influence the compression ratio. To present the rendering speeds more effectively, we normalized all compression ratios relative to the corresponding base model's compression ratio. This visualization is showcased in \Cref{vnr:fig:fps-scaling}. Our tests spanned across four datasets, encompassing a wide spectrum of data sizes, and we observed near-constant scaling curves for all the datasets assessed. This outcome is anticipated, as the hash grid encoding method arranges parameters in a random-access array and concurrently computes hidden vectors at different encoding levels.}

\add{Finally}, we have integrated the CUDA inference kernel of \fvsrn directly into our rendering system and conducted a comprehensive performance comparison between \fvsrn and our implementation under identical conditions. Our results show that our in-shader approach outperforms \fvsrn's implementation for most datasets, even without the use of macro-cell acceleration; the only exception to this is the \dataset{10atmTemp} dataset. Our optimized sample streaming algorithm exhibits a significant performance advantage over \fvsrn, highlighting the algorithmic superiority of our approach. Moreover, the utilization of macro-cells further improves the rendering performance, enabling advanced illumination effects such as shadows and global illumination.

%% file: contents/conclusion.tex
\section{Conclusion}

We present a volumetric neural representation for scientific visualization that can be trained nearly instantaneously and achieve a compression ratio of 10-1000$\times$.
To render the representations, we develop a novel sample streaming rendering algorithm that can ray trace the volumetric neural representation at 10-60fps with full global illumination for the first time in the field. We also develop an out-of-core sampling method for extreme-scale data to create neural representations using data streaming techniques.
Finally, we demonstrate that it is possible to achieve interactive online training performance using our implementations.

This research direction  is still very much in its early stages, and there is much room for future work. 
First, our work outlines how online training can be achieved using the latest technology. The potential of online training in scientific visualization  has not been fully explored yet. We are confident that it is possible to apply this technique to many other similar problems. 
Second, our training and rendering methods do not yet support distributed data while data distribution is performed in most scientific simulations. Extending our methods to support distributed and multi-resolution data structures is an essential next step.
Third, even though our volumetric neural representation is designed to support direct volume rendering, developing similar techniques for other visualization methods, such as isosurfaces, is also possible. Techniques such as the deep signed distance function~\cite{park2019deepsdf} have already provided clues about constructing a surface representation using neural networks. Therefore, applying such techniques to typical scientific visualization workflows is promising.
Next, we still lack understanding of the knowledge learned by the hash-table encoding. However, the possibility of using it directly as a macro-cell grid or a latent representation of the original data is intriguing.
\add{Finally, in instances where the neural representation is unable to adequately fit the data due to its limited size, and expanding the network is not feasible, the adoption of more advanced optimization techniques, such as adaptive sampling, may prove beneficial. Thus, examining how these advanced optimization techniques influence data reconstruction quality also represents a compelling direction for future study.}

With this work, we have taken an essential step towards enabling real-time, memory-efficient volume visualization for terascale applications using volumetric neural representations. We hope this project will spawn subsequent research to drive the exascale evolution of data processing.

%% file: contents/acknowledgments.tex
This project is sponsored in part by the U.S. Department of Energy through grant DE-SC0019486.
The ammonia/hydrogen/nitrogen-air flames dataset is kindly provided by Dr. Martin Rieth. 
The channel flow simulation dataset is kindly provided by Dr. Myoungkyu Lee.

%% file: contents/7.supplementary.tex
\section{In-Shader Inference Details}

\begin{figure*}[tb]
    \centering
    \includegraphics[width=\linewidth]{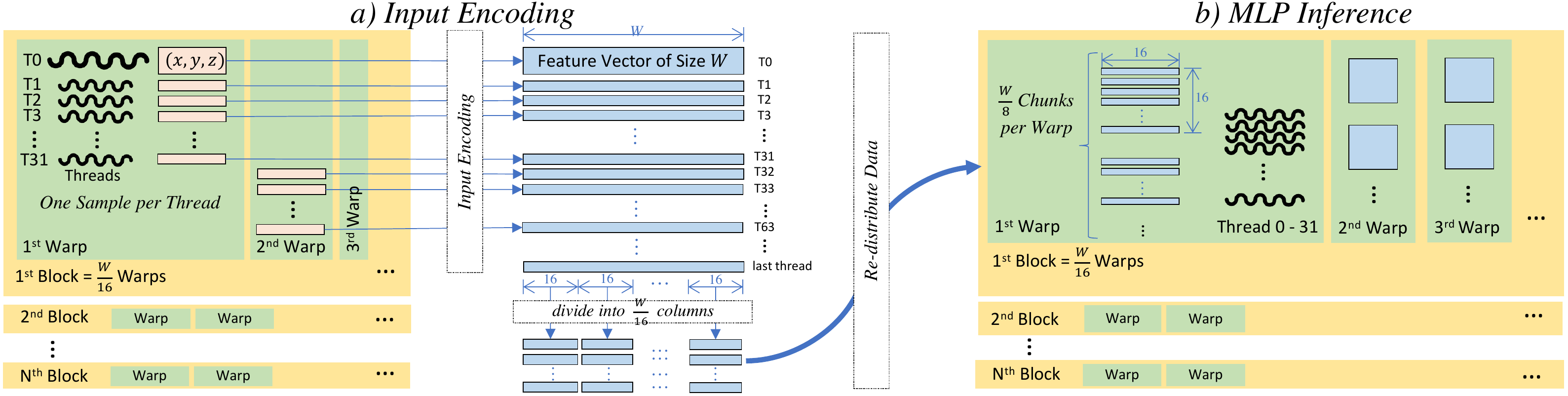}
    \caption{a) A typical GPU ray tracer produces one sample coordinate $(x,y,z)$ per thread. The input encoding is also computed by each thread independently, each producing a feature vector of size $W$. These feature vectors are then divided into $\frac{W}{16}$ 16-element-sized smaller vectors, written into the shared memory. b) The MLP inference is computed by re-distributing feature vectors between threads. Particularly, threads in a warp collaboratively compute the matrix multiplication of a $16\times16$ data chunk each time using a tensor core and iteratively computes $\frac{W}{8}$ chunks. There are $\frac{W}{16}$ warps in a block, computing $2W$ feature vectors in total.}
    \label{vnr:fig:in-shader}
\end{figure*}

The intuitive approach is to infer the neural network inside a shader program directly.
Thus we develop a fully customized network inference routine in the \tcnn framework using native CUDA.
Specifically, our routine involves the following adjustments.
First, we eliminate all the global GPU memory access and have the hash grid encoding layer fully fused into the MLP calculation. Inputs and outputs are managed locally within each threadblock and passed to the network via shared memory.
Second, as illustrated by~\Cref{vnr:fig:in-shader}, our routine breaks an MLP layer into 16-neuron-sized groups and allows the computation within each group to be completed by a 32-thread warp for 16 consecutive feature vectors. 
Thus, each warp only handles multiplications between two 16$\times$16 matrices, which can be accelerated using GPU tensor cores.
Then, all the $\frac{W}{16}$ groups are combined to form a GPU threadblock and executed in parallel.
As a result, each threadblock processes exactly $16$ inputs at a time.
In a typical GPU-based ray tracer, each GPU thread handles exactly one ray and only produces one input coordinate at a time.
We let each warp iterate for $\frac{W}{8}$ times to compensate for that.
Third, because now all the rendering kernel needs to be launched with a particular configuration, we also provide a templated CUDA wrapper function for launching kernels.
This wrapper function can take the ray-generation kernel as a callback and start the rendering with correct threadblock configurations and shared memory sizes.
Thus, we can easily integrate our in-shader inference method into a standard GPU volume ray tracing algorithm.
Finally, GPU tensor cores require all the threads in the calling warp to be active.
Thus, the control flow within a rendering kernel also should be carefully managed. In our implementations, we use a \texttt{block\_any} helper function to manage this.
This function evaluates true for all threads within a threadblock if any of the thread-local-predicates \texttt{v} evaluates true.
Notably, this method is also concurrently and independently developed by Weiss~\etal~\cite{weiss2021fast} but based on a completely different network architecture.

\section{Network Configurations}
In this section, we list all the network configurations used by experiments discussed in this paper. The specific documentation about each parameter can be found in the documentation page of \tcnn~\cite{tiny-cuda-nn}.

\subsection{Default Optimizer and Loss Function} 
Default optimizer and loss settings used in this paper, unless specified otherwise.
\begin{lstlisting}
"optimizer": {
    "otype": "ExponentialDecay",
    "decay_start": 2000,
    "decay_interval": 1000,
    "decay_base": 0.99,
    "nested": {
        "otype": "Adam",
        "learning_rate": 0.005,
        "beta1": 0.9,
        "beta2": 0.999,
        "epsilon": 1e-15,
        "l2_reg": 1e-06
    }
},
"loss": { "otype": "L1" }
\end{lstlisting}

\subsection{Ablation Study: Encoding Comparison} 
The network configurations used in \Cref{vnr:sec:ablation} and \Cref{vnr:fig:compare-encoding}.

\begin{lstlisting}
"encoding": {
    "otype": "Identity", // component type
    "scale": 1.0, // scaling of each dimension
    "offset": 0.0 // added to each dimension
},
"network": {
    // CutlassMLP is slower, but can support 
    "otype": "CutlassMLP", // more neurons
    "n_neurons": 256,
    "n_hidden_layers": 15
}
\end{lstlisting}

\begin{lstlisting}
"encoding": {
    "otype": "OneBlob",
    "n_bins": 64
},
"network": {
    "otype": "CutlassMLP",
    "n_neurons": 256,
    "n_hidden_layers": 14
}
\end{lstlisting}

\begin{lstlisting}
"encoding": {
    "otype": "Frequency",
    "n_frequencies": 32
},
"network": {
    "otype": "CutlassMLP",
    "n_neurons": 256,
    "n_hidden_layers": 14
}
\end{lstlisting}

\begin{lstlisting}
"encoding": {
    "otype": "DenseGrid",
    "n_levels": 2,
    "n_features_per_level": 2,
    "base_resolution": 37
},
"network": {
    "otype": "CutlassMLP",
    "n_neurons": 64,
    "n_hidden_layers": 4
}
\end{lstlisting}

\begin{lstlisting}
"encoding": {
    "otype": "HashGrid",
    "n_levels": 8,
    "n_features_per_level": 4,
    "log2_hashmap_size": 15,
    "base_resolution": 14
},
"network": {
    "otype": "CutlassMLP",
    "n_neurons": 64,
    "n_hidden_layers": 4
}
\end{lstlisting}

\subsection{Ablation Study: Hyperparameter Study} 
Two highlighted networks in the hyperparameter study and \Cref{vnr:fig:paramscan}. Hash grid encoding and fully-fused MLP were used for both networks. A simplified description scheme is used for conciseness.

\begin{lstlisting}
// large network
"params": {
    "n_levels": 16,
    "n_features_per_level": 8,
    "log2_hashmap_size": 19,
    "base_resolution": 4,
    "n_neurons": 64,
    "n_hidden_layers": 8
}
\end{lstlisting}
\begin{lstlisting}
// small network
"params": {
    "n_levels": 8,
    "n_features_per_level": 8,
    "log2_hashmap_size": 19,
    "base_resolution": 4,
    "n_neurons": 64,
    "n_hidden_layers": 4
}
\end{lstlisting}

\subsection{Our Networks used for \Cref{vnr:tab:quality}} 
\begin{lstlisting}
// rm_t60
"params": {
    "n_levels": 2,
    "n_features_per_level": 8,
    "log2_hashmap_size": 15,
    "base_resolution": 32,
    "n_neurons": 32,
    "n_hidden_layers": 3
}
\end{lstlisting}
\begin{lstlisting}
// vmhead
"params": {
    "n_levels": 2,
    "n_features_per_level": 8,
    "log2_hashmap_size": 15,
    "base_resolution": 32,
    "n_neurons": 32,
    "n_hidden_layers": 3
}
\end{lstlisting}

\begin{lstlisting}
// 1atmTemp
"params": {
    "n_levels": 10,
    "n_features_per_level": 8,
    "log2_hashmap_size": 19,
    "base_resolution": 4,
    "n_neurons": 64,
    "n_hidden_layers": 3
}
\end{lstlisting}
\begin{lstlisting}
// 10atmTemp
"params": {
    "n_levels": 10,
    "n_features_per_level": 8,
    "log2_hashmap_size": 19,
    "base_resolution": 4,
    "n_neurons": 64,
    "n_hidden_layers": 4
}
\end{lstlisting}

\begin{lstlisting}
// 1atmHR
"params": {
    "n_levels": 8,
    "n_features_per_level": 8,
    "log2_hashmap_size": 19,
    "base_resolution": 4,
    "n_neurons": 64,
    "n_hidden_layers": 4
}
\end{lstlisting}
\begin{lstlisting}
// 10atmHR
"params": {
    "n_levels": 8,
    "n_features_per_level": 8,
    "log2_hashmap_size": 19,
    "base_resolution": 4,
    "n_neurons": 64,
    "n_hidden_layers": 4
}
\end{lstlisting}

\begin{lstlisting}
// Chameleon
"params": {
    "n_levels": 8,
    "n_features_per_level": 8,
    "log2_hashmap_size": 19,
    "base_resolution": 4,
    "n_neurons": 64,
    "n_hidden_layers": 3
}
\end{lstlisting}

\begin{lstlisting}
// MechHand
"params": {
    "n_levels": 11,
    "n_features_per_level": 8,
    "log2_hashmap_size": 17,
    "base_resolution": 4,
    "n_neurons": 64,
    "n_hidden_layers": 5
}
\end{lstlisting}

\begin{lstlisting}
// SuperNova
"params": {
    "n_levels": 8,
    "n_features_per_level": 8,
    "log2_hashmap_size": 17,
    "base_resolution": 4,
    "n_neurons": 64,
    "n_hidden_layers": 2
}
\end{lstlisting}

\begin{lstlisting}
// Instability
"params": {
    "n_levels": 11,
    "n_features_per_level": 8,
    "log2_hashmap_size": 19,
    "base_resolution": 4,
    "n_neurons": 64,
    "n_hidden_layers": 5
}
\end{lstlisting}

\begin{lstlisting}
// PigHeart
"params": {
    "n_levels": 6,
    "n_features_per_level": 4,
    "log2_hashmap_size": 16,
    "base_resolution": 4,
    "n_neurons": 16,
    "n_hidden_layers": 2
}
\end{lstlisting}

\begin{lstlisting}
// PigHeart (LDR)
"params": {
    "n_levels": 9,
    "n_features_per_level": 8,
    "log2_hashmap_size": 19,
    "base_resolution": 4,
    "n_neurons": 64,
    "n_hidden_layers": 5
}
\end{lstlisting}

\begin{lstlisting}
// DNS (hand-picked)
"loss": {
    "otype": "L2"
},
"params": {
    "n_levels": 16,
    "n_features_per_level": 4,
    "log2_hashmap_size": 24,
    "base_resolution": 16,
    "n_neurons": 64,
    "n_hidden_layers": 8
}
\end{lstlisting}

\subsection{\fvsrn Networks used for \Cref{vnr:tab:quality}} 

\begin{lstlisting}
// HYBRID TRAINING
python volnet/train_volnet_lite.py $data.json     \
   --train:mode world                             \
   --train:samples 256**3                         \
   --train:sampler_importance 0.01                \
   --train:batchsize 64*64*128                    \
   --rebuild_dataset 51                           \
   --val:copy_and_split                           \
   --outputmode density:direct                    \
   --lossmode density                             \
   --layers ${layers}                             \
   --activation SnakeAlt:1                        \
   --fouriercount ${fouriercount}                 \
   --fourierstd -1                                \
   --volumetric_features_resolution ${latent_res} \
   --volumetric_features_channels 16              \
   -l1 1                                          \
   -lr 0.01                                       \
   --lr_step 120                                  \
   -i 200 

// mechhand, No. params fV-SRN=3473665, OURS=3462656
layers=64:64:64:64
latent_res=60
fouriercount=30

// chameleon, No. params fV-SRN=15072577, OURS=14992896
layers=64:64:64
latent_res=98 
fouriercount=30

// supernova, No. params fV-SRN=4609281, OURS=4503040
layers=64:64
latent_res=66 
fouriercount=30

// 1atmHR, No. params fV-SRN=15076737, OURS=14996992
layers=64:64:64:64
latent_res=98 
fouriercount=30

// 10atmHR, No. params fV-SRN=15076737, OURS=14996992
layers=64:64:64:64
latent_res=98 
fouriercount=30

// 1atmTemp, No. params fV-SRN=23718209, OURS=23382528
layers=64:64:64
latent_res=114
fouriercount=30

// 10atmTemp, No. params fV-SNR=23722369, OURS=23386624
layers=64:64:64:64
latent_res=114
fouriercount=30

// rm_t60 No. fV-SRN=527969, INR=527360
layers=32:32:32
latent_res=32
fouriercount=14

// vmhead
layers=32:32:32
latent_res=32
fouriercount=14

\end{lstlisting}

\subsection{Networks used for \Cref{vnr:tab:training-extra}} 
\begin{lstlisting}
// our method
"params": {
    "n_levels": 8,
    "n_features_per_level": 8,
    "log2_hashmap_size": 19,
    "base_resolution": 4,
    "n_neurons": 64,
    "n_hidden_layers": 4
}
\end{lstlisting}

\begin{lstlisting}
// reference: https://github.com/matthewberger/neurcomp
"neurcomp": {
    "grad_lambda": 0, 
    "n_layers": 8,
    "layers": [ 256, 256, 256, 256, 256, 256, 256, 256 ],
    "batchSize": 20480, 
    "oversample": 16,
    "n_passes": 75, 
    "pass_decay": 10,
    "lr": 1e-05, 
    "lr_decay": 0.45,
    "is_residual": true,
    "is_cuda": true
}
\end{lstlisting}

\subsection{Compare with Down-Sampling (\Cref{vnr:fig:downsample-vs-inshader})}  
\begin{lstlisting}
"params": {
    "n_levels": 16,
    "n_features_per_level": 8,
    "log2_hashmap_size": 19,
    "base_resolution": 4,
    "n_neurons": 64,
    "n_hidden_layers": 4
}
\end{lstlisting}

\subsection{Sampling Pattern Study (\Cref{vnr:fig:training-methods})} 
\begin{lstlisting}
// 10atmTemp & Instability
"params": {
    "n_levels": 16,
    "n_features_per_level": 8,
    "log2_hashmap_size": 19,
    "base_resolution": 4,
    "n_neurons": 64,
    "n_hidden_layers": 4
}
\end{lstlisting}

\begin{lstlisting}
// DNS
"loss": {
    "otype": "L2"
},
"params": {
    "n_levels": 16,
    "n_features_per_level": 4,
    "log2_hashmap_size": 24,
    "base_resolution": 16,
    "n_neurons": 64,
    "n_hidden_layers": 8
}
\end{lstlisting}

\subsection{Macro-Cell Evaluation (\Cref{vnr:fig:mc-quality})}
\begin{lstlisting}
"params": {
    "n_levels": 16,
    "n_features_per_level": 8,
    "log2_hashmap_size": 19,
    "base_resolution": 4,
    "n_neurons": 64,
    "n_hidden_layers": 4
}
\end{lstlisting}

\section{Adaptivity with Time-Varying Data}

\begin{figure*}[tb]
    \centering
    \includegraphics[width=\linewidth]{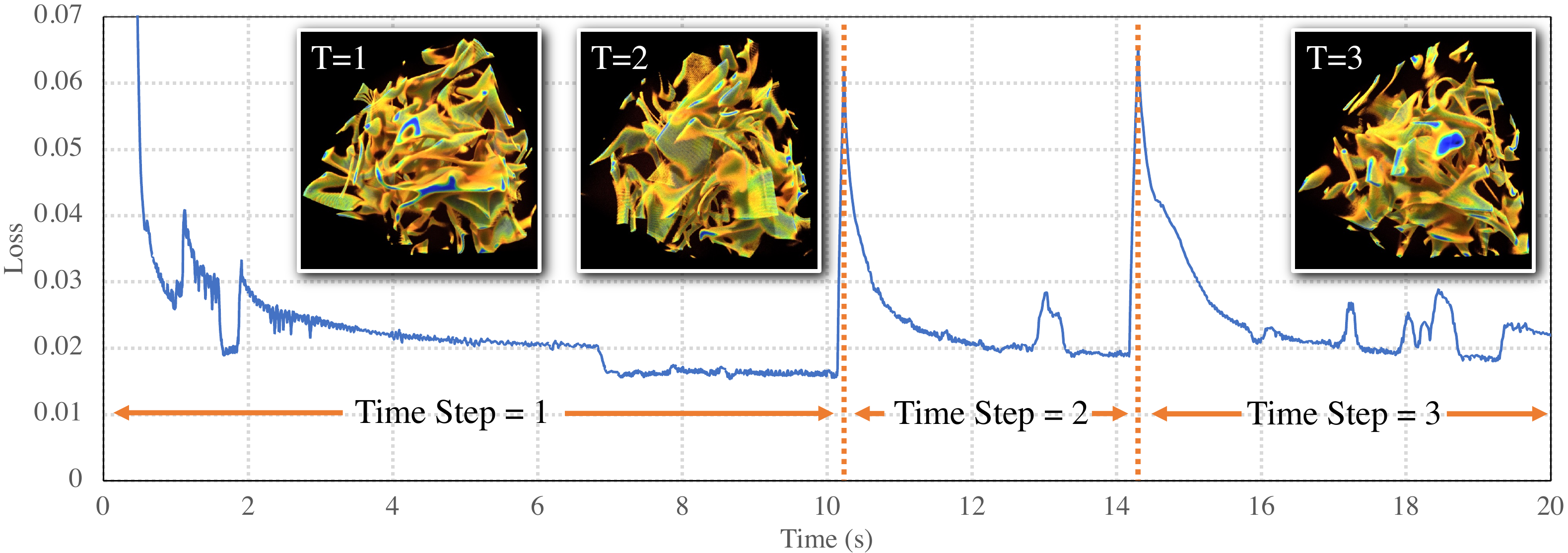}
    \caption{We can use a single volumetric neural representation to interactively visualize a time-varying dataset. In the experiment, the timestep was automatically incremented every 5 seconds.}
    \label{vnr:fig:time-varying-adaptivity}
\end{figure*}

Our neural representation method can also be applied to time-varying data directly with the help of our online-training capability. In \Cref{vnr:fig:time-varying-adaptivity} we demonstrate the real-time loss curve when interactively visualizing the multi-timestep vortices dataset provided by Deborah Silver at Rutgers University.
Because we want continuously train the same neural network (\textbf{without resetting network weights}) despite having a changeable target data, we did not train the network with learning rate decay in our experiment. Instead, we used a simple Adam optimizer with learning rate being fixed to $0.01$.
We changed the data timestep every 5 seconds.
We can see that our neural network could quickly adapt to the new timestep, and automatically adjust what has been learned within around two seconds.
Network used for \Cref{vnr:fig:time-varying-adaptivity} is listed below.
\begin{lstlisting}
"loss": { "otype": "L1" },
"optimizer": {
    "otype": "Adam",
    "learning_rate": 1e-4,
    "beta1": 0.9,
    "beta2": 0.999,
    "epsilon": 1e-8,
    "l2_reg": 1e-6
},
"encoding": {
    "otype": "HashGrid",
    "n_levels": 16,
    "n_features_per_level": 8,
    "log2_hashmap_size": 19,
    "base_resolution": 4
},
"network": {
    "otype": "FullyFusedMLP",
    "activation": "ReLU",
    "n_neurons": 64,
    "n_hidden_layers": 4,
    "output_activation": "ReLU"
}
\end{lstlisting}

\begin{figure}[tb]
    \centering
    \includegraphics[width=0.9\linewidth]{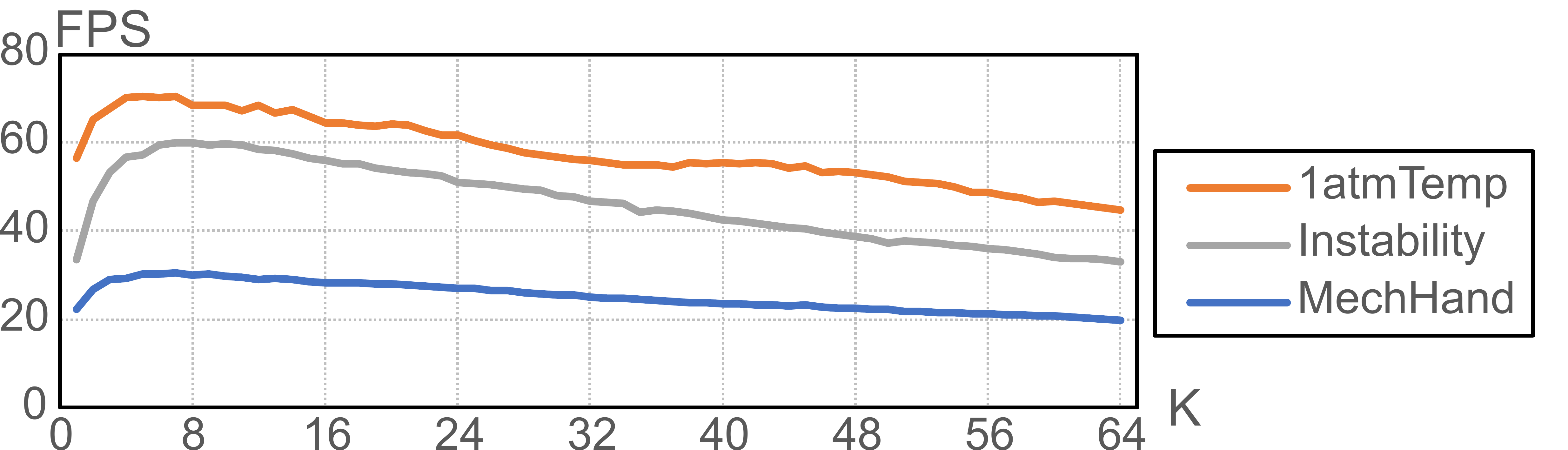}
    \vspace{-0.8em}
    \caption{In our ray marching sample-streaming renderer, a ray generates $K$ samples in each iteration. 
    We tested our algorithm with different $K$ values. The algorithm performs best when $K$ was around 8. The networks used here are the same as in \Cref{vnr:tab:quality}.}
    \label{vnr:fig:sample-streaming}
\end{figure}

\section{Tuning Our Sample-Streaming Algorithm}

As mentioned in \Cref{vnr:sec:sample-streaming-rm}, for the sample-streaming ray marching algorithm, the rendering performance can be improved by allowing a ray to take more than one sample in each iteration ($K$$>$1). In \Cref{vnr:fig:sample-streaming} we show the experiment to tune $K$. We can see that the algorithm performs best when $K$ was around 8, which is the value we used in this paper.

\section{Compare Our Sample-Streaming Algorithm with the Related Work}

Notably, M{\"u}ller~\etal~\cite{muller2022instant} recently also proposed a volumetric path tracer involving implicit neural representations. We want to clarify the difference between their method and our sample streaming implementation.
In their method, a radiance and density field is directly fitted using the noisy output of a volumetric path tracer. Then, a standard NeRF ray marcher is used to render the fitted neural representation. Because the in-scattering terms are learned already, the ray-marched final image can match the result of an unbiased volumetric path tracer.
However, in their algorithm, the ground truth volume density field is always accessible; thus, the path tracer can scatter rays freely without leaving the render kernel. Later, their ray marcher also uses the ground truth density field for empty space skipping, as the neural representation has not been trained on these empty regions.
In our case, the ground truth volume density is not used. Therefore, we base all of our calculations on the neural representation. This includes empty space skipping, shadows, multi-scattering, and direct lighting, which makes our implementation independent from the ground truth and significantly more compact (and more sophisticated).

\begin{table}[tb]
\caption{Training times for \neurcomp networks.}
\label{vnr:tab:neurcomp}
\scriptsize\centering
\begin{tabu}{r|rc}
\toprule 
& \multicolumn{2}{c}{\makecell{\neurcomp\\({3.6MB})}}\\
Dataset       &Time            & PSNR   \\\midrule
{1atmTemp   } & 231\textit{m}  & 33.4   \\
{10atmTemp  } & 308\textit{m}  & 30.1   \\
{1atmHR     } & 403\textit{m}  & 28.8   \\
{10atmHR    } & 543\textit{m}  & 27.0   \\
{Chameleon  } & 1458\textit{m} & 50.1   \\
{MechHand   } & 23\textit{m}   & 46.2   \\
{SuperNova  } & 58\textit{m}   & 45.3   \\\bottomrule
\end{tabu}
\end{table}

\section{Additional Results and Images}

\begin{figure*}[tb]
    \centering
    \includegraphics[width=\linewidth]{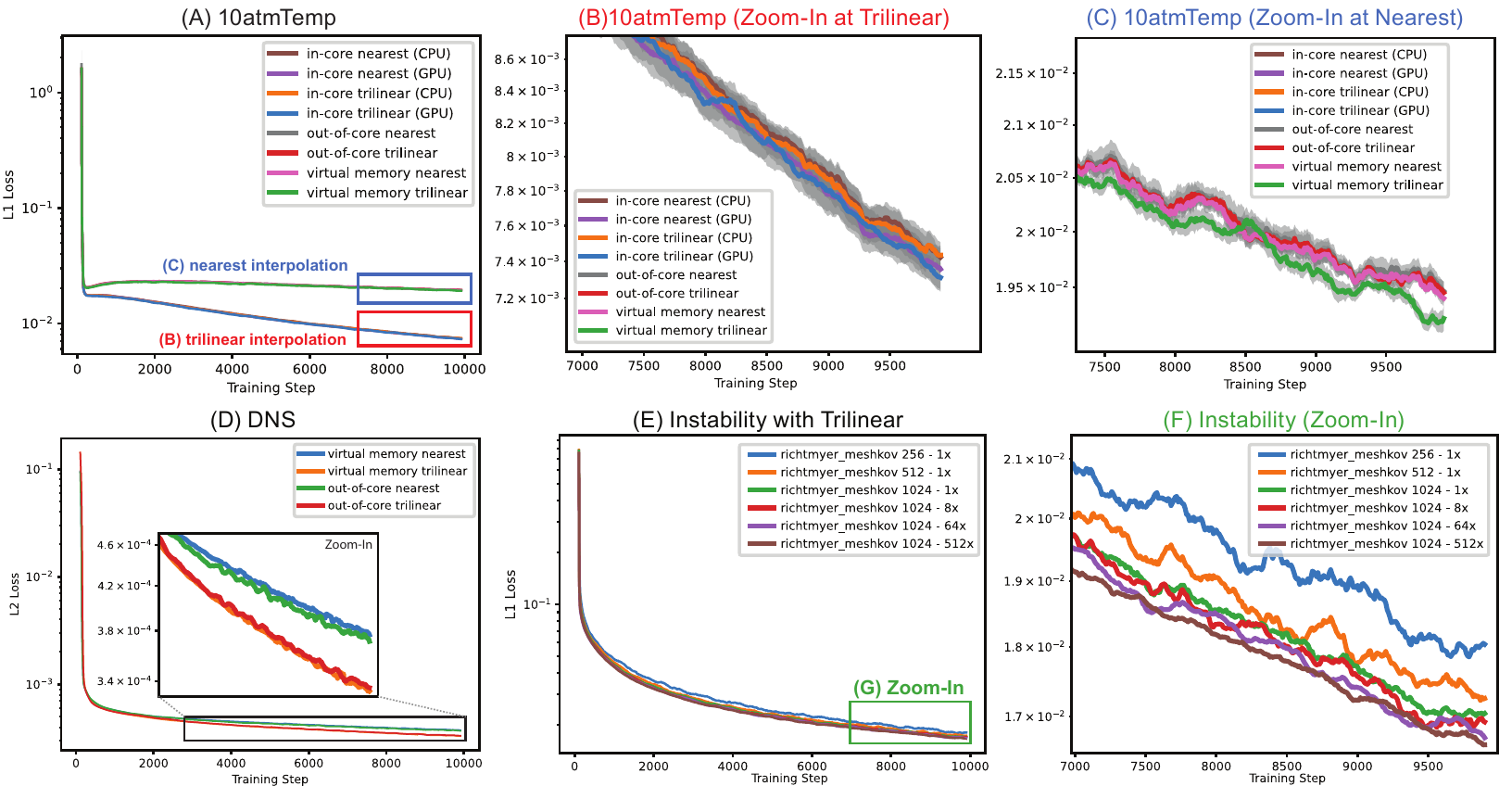}
    \caption{Larger version of \Cref{vnr:fig:training-methods}.}
    \label{vnr:fig:training-methods-large}
\end{figure*}

\begin{figure*}[tb]
    \centering
    \includegraphics[width=\linewidth]{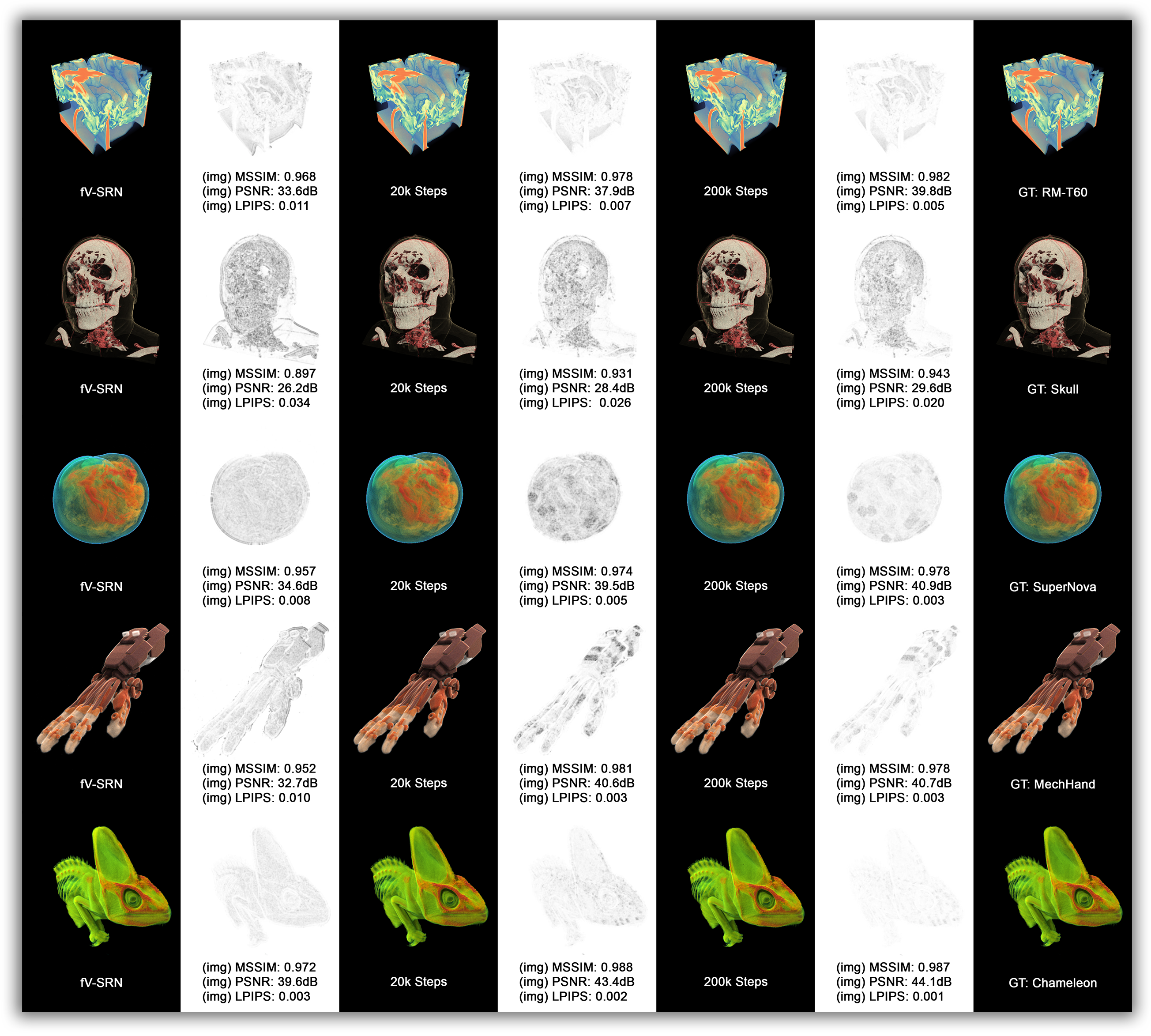}
    \caption{A larger version of \Cref{vnr:fig:renderings}.}
    \label{vnr:fig:renderings-additional-0}
\end{figure*}

\begin{figure*}[tb]
    \centering
    \includegraphics[width=\linewidth]{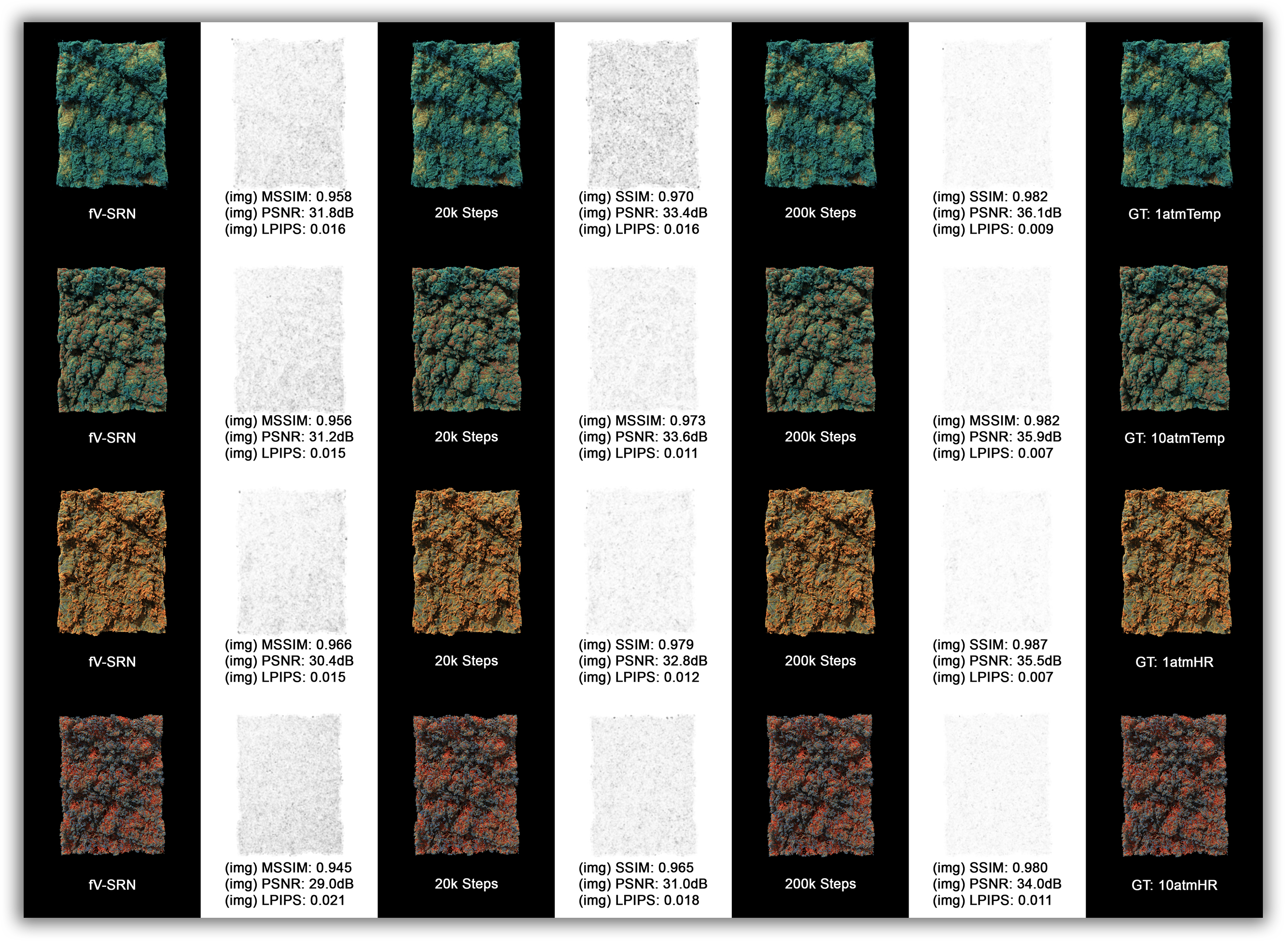}
    \caption{Additional rendering quality comparisons for datasets that are absent from \Cref{vnr:fig:renderings}.}
    \label{vnr:fig:renderings-additional-1}
\end{figure*}

\begin{figure*}[b]
    \centering
    \includegraphics[width=0.714\linewidth]{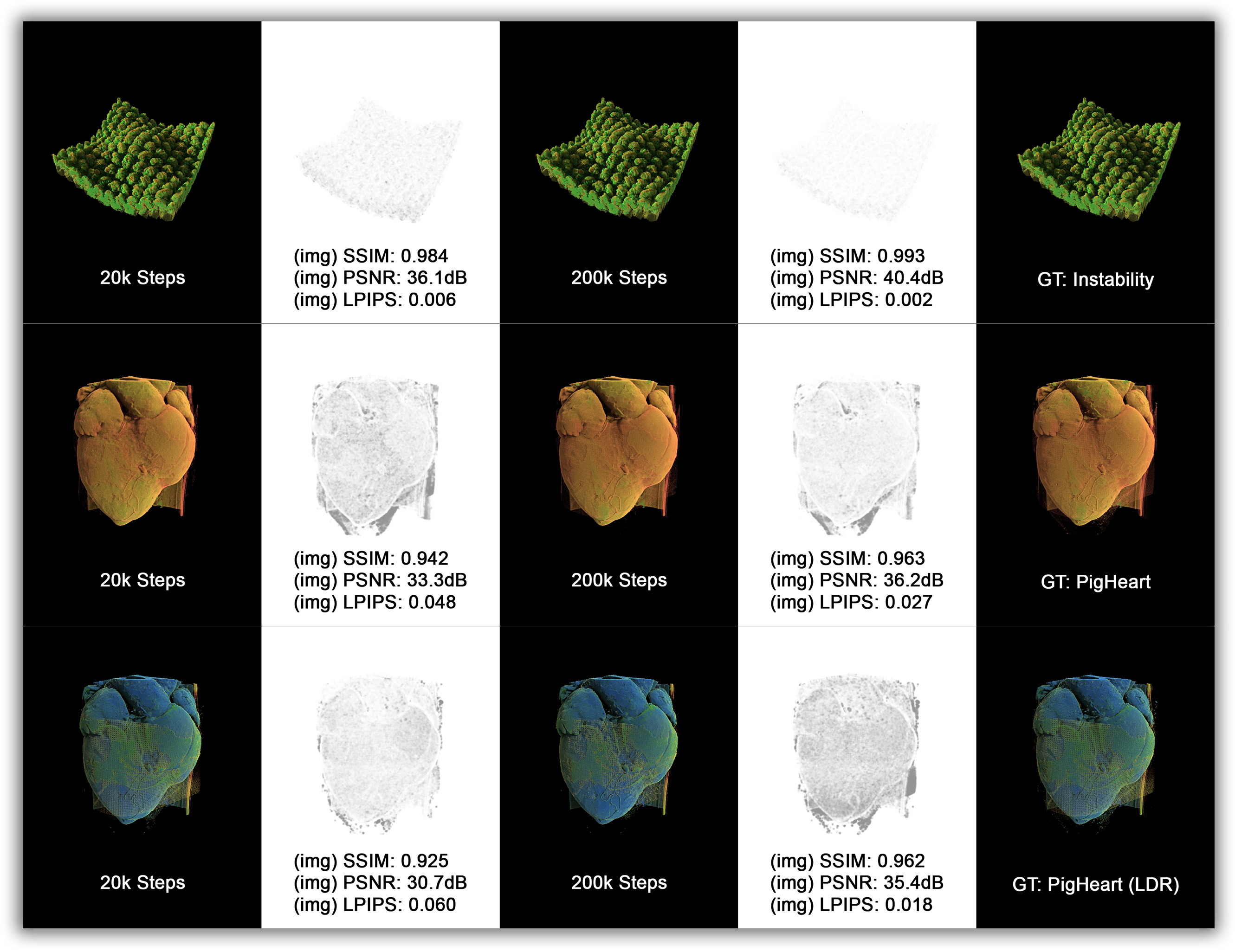}
    \caption{Additional rendering quality comparisons for datasets that are absent from \Cref{vnr:fig:renderings}.}
    \label{vnr:fig:renderings-additional-2}
\end{figure*}

Here we provide additional rendering results that are not shown in the paper. 
\label{vnr:tab:neurcomp} reports \neurcomp networks' training times. These networks are evaluated in \Cref{vnr:tab:training-extra}.
Rendering quality comparisons of datasets that are not shown in \Cref{vnr:fig:renderings} can be found in
\Cref{vnr:fig:renderings-additional-0}, \Cref{vnr:fig:renderings-additional-1}, and 
\Cref{vnr:fig:renderings-additional-2}.
Next, a larger version of \Cref{vnr:fig:training-methods} can be found in \Cref{vnr:fig:training-methods-large}.
\add{Finally, we provide further details pertaining to the experiment outlined in \Cref{vnr:fig:fps-scaling}.}

\begin{table}[tb]
\caption{\add{We offer supplementary information regarding the experiment depicted in \Cref{vnr:fig:fps-scaling}. Here, NCR denotes the Normalized Compression Ratio.}}
\label{vnr:tab:fps-scaling-details}
\scriptsize\centering
\begin{tabu}{r|rrrr}
\toprule
& \multicolumn{4}{c}{Value of \texttt{log2\_hashmap\_size}}\\
Dataset    & Base (NCR) & 1st (NCR) & 2nd (NCR) & 3rd (NCR) \\\midrule
{PigHeart} & 16 (1$\times$) & 15 (1.6$\times$) & 14 (3.1$\times$) & 13 (5.6$\times$) \\
{10atmTemp}& 19 (1$\times$) & 18 (1.8$\times$) & 17 (3.5$\times$) & 16 (6.8$\times$) \\
{Chameleon}& 19 (1$\times$) & 18 (1.7$\times$) & 17 (3.3$\times$) & 16 (6.6$\times$) \\
{SuperNova}& 17 (1$\times$) & 16 (1.9$\times$) & 15 (3.3$\times$) & 14 (6.4$\times$) \\
\bottomrule
\end{tabu}
\end{table}